\documentclass[twocolumn,nofootinbib,prd,aps,tightenlines]{revtex4}

\usepackage{slashed}
\usepackage{braket}
\usepackage{graphicx}

{\count255=\time\divide\count255 by 60 \xdef\hourmin{\number\count255}
  \multiply\count255 by-60\advance\count255 by\time
  \xdef\hourmin{\hourmin:\ifnum\count255<10 0\fi\the\count255}}

\newcommand{\nn}{\nonumber \\ }
\newcommand\eUV{\epsilon_{\text{UV}}}
\newcommand\eIR{\epsilon_{\text{IR}}}

\def\rgel{\mathsf{L}}

\def\lQ{\mathsf{L}_{Q}}
\def\lM{\mathsf{L}_{M}}
\def\lchi{\mathsf{L}_{\chi}}
\def\lm{\mathsf{L}_{m}}
\def\lmi{\mathsf{L}_{m_1}}
\def\lmii{\mathsf{L}_{m_2}}
\def\lni{\mathsf{L}_{1}}
\def\lnii{\mathsf{L}_{2}}
\def\lnm{\mathsf{L}_{-}}

\def\darr#1{\raise1.5ex\hbox{$\leftrightarrow$}\mkern-16.5mu #1}
\def\rd{{{\rm d}}}
\def\aem{\alpha_{\text{em}}}

\def\darr#1{\raise1.5ex\hbox{$\leftrightarrow$}\mkern-16.5mu #1}

\begin{document}

\title{Electroweak Corrections in High Energy Processes using Effective Field Theory}

\author{Jui-yu Chiu}
\affiliation{Department of Physics, University of California at San Diego,
  La Jolla, CA 92093}

\author{Frank Golf}
\affiliation{Department of Physics, University of California at San Diego,
  La Jolla, CA 92093}

\author{Randall Kelley}
\affiliation{Department of Physics, University of California at San Diego,
  La Jolla, CA 92093}

\author{Aneesh V.~Manohar}
\affiliation{Department of Physics, University of California at San Diego,
  La Jolla, CA 92093}
\date{\today\quad\hourmin}

\begin{abstract}
Electroweak Sudakov logarithms at high energy, of the form $(\alpha/\sin^2\theta_W)^n\log^m s/M_{Z,W}^2$, are summed using effective theory methods. The corrections are computed to processes involving two external particles in the standard model. The results include non-zero particle masses, such as the $t$-quark mass, electroweak mixing effects which lead to unequal $W$ and $Z$ masses, and radiative Higgs corrections proportional to the Yukawa couplings. We show that the matching at the scale $M_{W,Z}$ has a term at most linear in $\log s/\mu^2$ to all orders. The effective theory formalism is compared with, and extends, previous work based on infrared evolution equations.
\end{abstract}

\maketitle
\tableofcontents

\section{Introduction}
\label{I}

The Large Hadron Collider (LHC) has a center-of-mass energy of $\sqrt s =14$~TeV, and will be able to measure collisions with a partonic center-of-mass energy of several TeV, more than an order of magnitude larger than the masses of the electroweak gauge bosons. Radiative corrections to scattering processes depend on the ratio of mass scales, and radiative corrections at high energy depend on logarithms of the form $\log s/M^2_{W,Z}$.  In high energy exclusive processes, radiative corrections are enhanced by two powers of a large logarithm for each order in perturbation theory, and the logarithms are often referred to as Sudakov (double) logarithms.  Electroweak Sudakov corrections are not small at LHC energies, since $\alpha \log^2 s/M^2_{W,Z}/(4 \pi \sin^2 \theta_W) \sim 0.15$ at $\sqrt{s}=4$~TeV. These Sudakov corrections lead to a breakdown of fixed order perturbation theory, and need to be summed to all orders.

Electroweak corrections at high energy have double logarithms, even for processes which are conventionally called inclusive, such as the total $e^+e^-$ cross-section at large angles, because the colliding particles are not electroweak gauge singlets~\cite{ccc}. There are no electroweak singlet fields in the standard model.  A composite particle such as the proton, while a color singlet, is not an electroweak singlet.

There is an extensive literature on electroweak Sudakov effects~\cite{ciafaloni,fadin,kps,fkps,jkps,jkps4,beccaria,dp1,dp2,hori,beenakker,dmp,pozzorini,js}. The computations use infrared evolution equations~\cite{fadin}, based on an analysis of the infrared structure of the perturbation theory amplitude and a factorization theorem for the Sudakov form factor~\cite{pqcd}. These summations have been  checked against one-loop~\cite{beccaria,dp1,dp2} and two-loop~\cite{hori,beenakker,dmp,pozzorini,js} computations.

The Sudakov logarithm $\log(s/M_{W,Z}^2)$ can be thought of as an infrared logarithm in the electroweak theory, since it diverges as $M_{W,Z}\to0$. By using an effective field theory (EFT), these infrared logarithms in the original theory can be converted to ultraviolet logarithms in the effective theory, and summed using standard renormalization group techniques. The effective theory needed is soft-collinear effective theory (SCET)~\cite{BFL,SCET}, which has been used to study high energy processes in QCD, and to perform Sudakov resummations arising from radiative gluon corrections.

This paper studies high energy electroweak Sudakov corrections using SCET, and expands on our previous work~\cite{cgkm1}.  In Ref.~\cite{cgkm1}, we showed how to compute $\log s/M^2_{W,Z}$ corrections to the Sudakov form factor for massless fermions using EFT methods. In this paper, the results are generalized to massive fermions such as the top quark, and include radiative corrections due to Higgs exchange.
The corrections are computed  without assuming that the Higgs and electroweak gauge bosons are degenerate in mass, as in previous calculations.  A new feature of EFT matching, the existence of single logarithmic matching corrections~\cite{cgkm1}, is discussed in detail, and proven to be true to all orders in perturbation theory. This paper discusses the Sudakov form factor computation in detail. The Sudakov form factor is not of direct relevance to LHC processes, but it allows us to illustrate the EFT method for operators involving two external particles.  The computations of the Sudakov form factor given in this paper can be used to compute electroweak corrections to processes relevant for the LHC, such as dijet production, $t \bar t$ production, or squark pair production, which involve operators with four external particles. The results are given in a future publication~\cite{cgkm3}, and can be obtained from the computations given in this paper by summing over all pairs of external particles with the appropriate group theoretic factors.

The outline of the calculation is given in Sec.~\ref{sec:outline}. The SCET formalism and the full theory we use for our calculations are described in Sec.~\ref{sec:scet}. Known results on the exponentiation of the Sudakov form factor, and a comparison of the infrared evolution equation formalism with the SCET approach is given in Sec.~\ref{sec:expon}. Section~\ref{sec:massless} discusses the calculation of Sudakov corrections for massive gauge bosons and massless external particles. Section~\ref{sec:linear} gives the proof that there is at most a single logarithm found in the matching condition to all orders in perturbation theory, and consistency conditions on the matching coefficients and anomalous dimensions are given in Sec.~\ref{sec:cc}. The extension to massive external particles is given in Sec.~\ref{sec:mass} for all possible hierarchies of mass-scales, including cases in which particle masses are not widely separated, so that multiple scales have to be integrated out simultaneously. Massive scalar exchange graphs, relevant for Higgs exchange, are computed in Sec.~\ref{sec:scalar}. Applications of the formalism to electroweak Sudakov corrections in the standard model is given in Sec.~\ref{sec:stdmodel} for light quarks, the top quark, and leptons.

{\bf Notation:} We use  $a(\mu)\equiv\alpha(\mu)/(4\pi)$, and $a_i(\mu)\equiv\alpha_i(\mu)/(4\pi)$ where $i=s,2,1$ for the QCD, $SU(2)$ and $U(1)$ couplings in the standard model. Hypercharge is normalized so that $Q=T_3+Y$. Logarithms are denoted by $\mathsf{L}_A \equiv\log A^2/\mu^2$, for $A=Q,M,m_1,m_2$. $C_F$ and $T_F$ are the Casimir and index for the external particles. We use the subscript $F$ for both fermions and scalars, to avoid rewriting the same expression twice.
 
\section{Outline of Calculation}
\label{sec:outline}

The physical quantity we study is the Sudakov form factor in the Euclidean region, defined as the amplitude $F_E(Q^2)=\braket{p_2|\mathcal{O} |p_1}$ for the scattering of on-shell particles $p_i^2=m_i^2$ by an operator $\mathcal{O}$, with $Q^2=-(p_2-p_1)^2>0$.  The timelike Sudakov form factor is given by analytic continuation, $F(s) = F_E(-s-i0^+)$, so that $\log(Q^2/\mu^2) \to \log(s/\mu^2)-i\pi$.  

We will compute $F_E(Q^2)$ for fermion scattering by $\mathcal{O}=\bar\psi\gamma^\mu\psi,\, \bar\psi \psi,\, \bar\psi\sigma^{\mu\nu}\psi$, scalars scattering by $\mathcal{O}=\phi^\dagger \phi,\, i(\phi^\dagger D^\mu \phi-D^\mu\phi^\dagger  \phi)$, and fermion to scalar (or vice-versa) scattering by $\mathcal{O}=\bar\psi \phi$.   All operators are taken to be gauge singlets so the incoming and outgoing particles have the same gauge quantum numbers, but not necessarily the same mass.

The form factor, $F_E(Q^2)$ is computed using a sequence of effective theories.  
For the high energy process considered is this paper, there are several widely separated scales and we must switch to the relevant theory as we move between scales.  At scales higher than $Q^2$, the theory is the original gauge theory, referred to as the full theory in EFT terminology. The precise theory, and the SCET formalism used are given in Sec.~\ref{sec:scet}.

As we move to scales below $Q^2$ we transition to an effective field theory (SCET) where degrees of freedom with offshellness on the order of  $Q^2$ are integrated out.  The full and EFT have the same infrared (IR) physics but different ultraviolet (UV) behavior and to ensure that the operators in the respective theories  have the same on-shell matrix elements, we must introduce a matching coefficient, $\exp[C(\mu)]$.  For later convenience, the matching coefficient is written as an exponential.
If the full theory is matched onto SCET at $\mu_Q$ then the matching coefficient is chosen so that 
\begin{eqnarray}
\braket{p_2|\mathcal{O}(\mu_Q)|p_1} &=& \exp[ C(\mu_Q) ]\braket{p_2|\widetilde{\mathcal{O}}(\mu_Q)|p_1} 
\end{eqnarray} 
where $\tilde{ \mathcal{O} }(\mu)$ is the EFT operator corresponding to the full theory operator $\mathcal{O}(\mu)$. The matching coefficient $\exp C(\mu_Q)$ is independent of infrared physics, and can be computed if perturbation theory is valid at $\mu_Q$. 
In general, a single operator $\mathcal{O}$ can match onto a set of operators $\widetilde\mathcal{O}_i$ in the EFT with the same quantum numbers.
This occurs, for example, for four-fermion operators in the analysis of high-energy parton scattering, and can be included by treating all the equations below as matrix equations, as is familiar from the well-known analysis of operator mixing. The matching coefficient $C(\mu_Q)$  contains $\log \mu_Q^2/Q^2$ terms, and there are no large logarithms if $\mu_Q$ is chosen to be of order $Q$. We will choose $\mu_Q=Q$, though any value of order $Q$ is acceptable. Any physical observable is independent of the choice for $\mu_Q$. It is conventional to choose $c(\mu)$, the coefficient of $\mathcal{O}$  in the full theory, to equal unity at $\mu=Q$. With this choice, which gives the usual normalization for $F_E(Q^2)$, $c(Q)=\exp C(Q)$ is the coefficient of $\widetilde\mathcal{O}$ in SCET at $\mu=Q$.
The evolution of $c(\mu)$ between scales is given by the renormalization group equation
\begin{equation}
\mu \frac{dc(\mu)}{d\mu} = \gamma(\mu)c(\mu),
\end{equation}
where $\gamma(\mu)$ is the anomalous dimension of $\widetilde{ \mathcal{O} }$ in the EFT.  

We must repeat this sequence of matching and renormalization group evolution as various energy scales are crossed, and more and more degrees of freedom are integrated out.  An advantage of the EFT approach is that it divides the full multiscale computation into several simpler pieces, each of which depends on a single scale. This allows one to easily identify which quantities are universal, and which ones depend on the specific process. In an EFT calculation, the IR divergences in the theory above a matching scale must match with the UV divergences in the theory below the matching scale. We have checked this explicitly for all the computations in this paper. In most of the tables, we have given only the finite parts of the graphs.

\section{SCET Formalism}
\label{sec:scet}

SCET is an effective theory that describes energetic particles, with energy of order $Q$, where $Q$ is some large scale which characterizes the scattering process.  SCET contains all the modes of the full theory with invariant mass much smaller than $Q^2$.  
The SCET fields and Lagrangian depend on two null four-vectors $n$ and $\bar n$,
with $n=(1,\bf{n})$ and $\bar n=(1,-\bf{n})$, where $\bf{n}$ is a unit vector, so that $\bar n \cdot n=2$. In the Sudakov problem, one works in the Breit frame, with $n$ chosen to be along the $p_2$ direction, so that $\bar n$ is along the $p_1$ direction. The momentum transfer $q$ has no time component, $q^0=0$, so that the particle is back-scattered. The light-cone components of a four-vector $p$ are defined by $p^+ \equiv n \cdot p$, $p^- \equiv \bar n \cdot p$.  In our problem, $p_1^-=p_{1\perp}=p_2^+=p_{2\perp}=0$, and $Q^2=p_1^+ p_2^-$.  A fermion moving in a direction close to $n$ is described by the $n$-collinear SCET field $\xi_{n,p}(x)$, where $p$ is a label momentum, and has components $\bar n \cdot p$ and $p_\perp$~\cite{BFL,SCET}.  It describes particles (on- or off-shell) with energy $2E=\bar n \cdot p$, and $p^2 \ll Q^2$. The SCET power counting is $p^- \sim Q$, $p^+ \sim Q\lambda^2$, $p_\perp \sim Q \lambda$, where $\lambda \ll 1$ is the power counting parameter used for the EFT expansion. The total momentum of the field $\xi_{n,p}(x)$ is $p+k$, where $k$ is the residual momentum of order $Q \lambda^2$ contained in the Fourier transform of $x$. Note that the label momentum $p$ only contributes to the minus and $\perp$ components of the total momentum. 

The gauge field is represented by several distinct fields in the effective theory: $n$-collinear fields $A_{n,p}(x)$ and $\bar n$-collinear fields $A_{\bar n,p}(x)$ with labels, and ultrasoft fields $A(x)$ with no label, analogous to the soft and ultrasoft fields introduced in NRQCD~\cite{lmr}. The $n$-collinear field contains gluons with momentum near the $n$-direction, and momentum scaling $\bar n \cdot p \sim Q$, 
$n \cdot p \sim Q \lambda^2$, $p_\perp \sim Q \lambda$, and the $\bar n$-collinear fields contain gluons moving near the $\bar n$-direction, with momentum scaling
$n \cdot p \sim Q$,  $\bar n \cdot p \sim Q \lambda^2$, $p_\perp \sim Q \lambda$.
The ultrasoft field contains gluons with all momentum components scaling as $Q \lambda^2$. 

The EFT fermion field satisfies the constraint
\begin{eqnarray}
\frac{ \slashed{n} \slashed{\bar n}}{4} \xi_{n,p} &=& \xi_{n,p},
\label{constraint}
\end{eqnarray}
where
\begin{eqnarray}
P_n = \frac{ \slashed{n} \slashed{\bar n}}{4},\quad
P_{\bar n} = \frac{ \slashed{\bar n} \slashed{n}}{4},\quad
P_n+P_{\bar n}=1,
\end{eqnarray}
are projection operators. The leading order fermion Lagrangian is~\cite{SCET}
\begin{eqnarray}
\bar \xi_{n,p} \frac{ \slashed{\bar n}}{2}\left( i n \cdot D + \frac{p_\perp^2}{\bar n \cdot p} \right) \xi_{n,p} + \ldots\, ,
\label{ferm}
\end{eqnarray}
where $iD=i\partial + g A$ is the ultrasoft covariant derivative, and $\ldots$ denotes terms involving the collinear gauge field.
The fermion propagator is
\begin{eqnarray}
\frac{ \slashed{n}\ \bar n \cdot p}{2p^2}.
\end{eqnarray}
The effective theory knows about the large momentum scale $Q$ through the labels $\bar n \cdot p_2$ and $n \cdot p_1$ on the fields $\xi_{n,p_2}$ and $\xi_{\bar n, p_1}$ for the outgoing and incoming particles.  As a result, SCET anomalous dimensions can depend on $Q$.  However, there are no modes in SCET which couple $\bar n \cdot p_2$ to $n \cdot p_1$, so that SCET does not contain modes with off-shellness of order $Q^2$, which are present in the full theory.

We will also need to introduce SCET fields to describe energetic scalar particles, such as the Higgs boson. We will use $\Phi_{n,p}$ as the $n$-collinear field for a scalar particle moving in a direction close to $n$, analogous to $\xi_{n,p}$ for fermions. The field $\Phi_{n,p}$ is normalized the same way as the full-theory field $\phi$, and produces scalar particles with amplitude unity. The scalar kinetic energy term becomes
\begin{eqnarray}
D_\mu \phi^\dagger D^\mu \phi \to  \Phi_{n,p}^\dagger 
\left[\left(\bar n \cdot p\right)\left( in \cdot D\right)+p_\perp^2\right]\Phi_{n,p}
\end{eqnarray}
in the effective theory.  It is also convenient to use a redefined scalar field,
\begin{eqnarray}
\phi_{n,p} = \sqrt{\left(\bar n \cdot p\right)}\, \Phi_{n,p},
\end{eqnarray}
in terms of which the kinetic term becomes
\begin{eqnarray}
L &=&   \phi_{n,p}^\dagger \left( i n \cdot D + \frac{p_\perp^2}{\bar n \cdot p} \right) \phi_{n,p}
\end{eqnarray}
and has the same normalization as the fermion Lagrangian Eq.~(\ref{ferm}). The rescaled scalar propagator is now
\begin{eqnarray}
\frac{1}{p^2} &\to& \frac{\bar n \cdot p}{p^2}.
\end{eqnarray}
$\phi_{n,p}$ produces scalar particles moving in the $n$-direction with amplitude $\sqrt{\left(\bar n \cdot p\right)}$. 

The theory we consider is a $SU(2)$ spontaneously broken gauge theory, with a Higgs in the fundamental representation, where all gauge bosons have a common mass, $M$.    This is the theory used in many previous computations~\cite{kps,fkps,jkps,jkps4,js}, and allows us to compare with previous results.   It is convenient, as in Ref.~\cite{js}, to write the group theory factors using $C_F$, $C_A$, $T_F$ and $n_F$, where $2n_F$ is defined to be the number of weak doublets Weyl fermions.\footnote{This convention  for $n_F$ is used in Ref.~\cite{js}. Note that the results only hold for $C_A=2$, since for an $SU(N)$ group with $N>2$, a fundamental Higgs does not break the gauge symmetry completely.}  We will consider this theory with fermionic and scalar matter fields in arbitrary gauge representations, with the fermions assumed to be vector-like. These fields are the external particles in the operators $\mathcal{O}$. We will also need to consider graphs which are analogous to Higgs exchange graphs in the standard model. For this purpose, we will add a gauge singlet scalar field $\chi$, which couples to the fermions and scalars via gauge-invariant interactions,
\begin{eqnarray}
L_{\text{int}} &=& - h_{\psi,i} \chi \bar \psi_i \psi_i - h_{\phi,i} \chi \phi_i^\dagger \phi_i,
\label{yuk}
\end{eqnarray}
$h_{\psi,i}$ is dimensionless, and $h_{\phi,i}$ has dimensions of mass. We will assume that $h_{\phi,i}$ is independent of $Q$ for power counting purposes. In our toy example, $\chi$ is a gauge singlet field, and does not break the gauge symmetry. The fermion masses are independent of the Yukawa couplings of $\chi$. The toy example Higgs field is a doublet, and breaks the gauge symmetry, but does not couple to the matter fields. In the standard model, the Higgs field breaks the gauge symmetry, and also has Yukawa couplings which generate fermion masses. 

\begin{figure}
\begin{center}
\includegraphics[width=4cm]{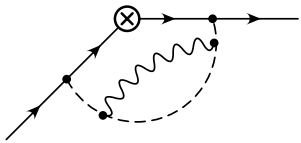}
\end{center}
\caption{\label{fig:absent} Graphs in the standard model involving both Yukawa and gauge couplings which have no analog in the toy example.}
\end{figure}

The computations are extended to the $SU(3) \times SU(2)_L\times U(1)_Y$ standard model in Sec.~\ref{sec:stdmodel}, including Higgs exchange corrections and unequal gauge boson masses. Our results are given to leading order in the EFT power counting, i.e.\ we neglect power corrections of the form $m_i^2/Q^2$, and $M^2/Q^2$, while retaining all logarithmic corrections $\log m_i^2/Q^2$ and $\log M^2/Q^2$. The gauge boson exchange graphs can be obtained from those of the toy model. In the standard model, the Higgs field breaks the gauge symmetry, and also has Yukawa couplings to the chiral fermions. At one-loop, we can obtain the Higgs exchange corrections from the $\chi$-exchange graphs in our toy example. Graphs with Higgs bosons coupling to both the fermions and the gauge bosons start at two-loops (see Fig.~\ref{fig:absent}).

\section{Exponentiation}
\label{sec:expon}

We start by summarizing some known properties of the Sudakov form-factor~\cite{collins} for the vector current. We will see later how the same expressions can be rederived using renormalization group methods in SCET. The Euclidean form-factor $F_E(Q^2)$ has the expansion ($\rgel=\log(Q^2/M^2)$)
\begin{eqnarray}
F_E &=& 1 \nn
&&+ \alpha\left( k_{12}\rgel^2+k_{11}\rgel + k_{10}\right)\nn
&&+\alpha^2\left(k_{24}\rgel^4+k_{23}
\rgel^3+k_{22}\rgel^2+k_{21}\rgel+k_{20}\right)\nn
&&+\alpha^3\left(k_{36}\rgel^6 + \ldots \right)+\ldots\, ,
\label{1}
\end{eqnarray}
with the $\alpha^n$ term having powers of $\rgel$ up to $\rgel^{2n}$. In the literature, the highest power of $\rgel$ is called the $\text{LL}_{\text{F}}$ term, the next power is called the $\text{NLL}_{\text{F}}$ term, etc.\ We have included the subscript $F$ (for the form-factor) to distinguish it from the renormalization group counting described below.

The series for $\log F_E(Q^2)$ takes a simpler form
\begin{eqnarray}
\log F_E &=&  \alpha\left(\tilde k_{12}\rgel^2+\tilde k_{11}\rgel + \tilde k_{10}\right)\nn
&&+\alpha^2\left(\tilde k_{23}
\rgel^3+ \tilde k_{22}\rgel^2+ \tilde k_{21}\rgel+\tilde k_{20}\right)\nn
&&+\alpha^3\left(\tilde k_{34}\rgel^4 + \ldots \right)+\ldots\, ,
\label{2}
\end{eqnarray}
with the $\alpha^n$ term having powers of $\rgel$ up to $\rgel^{n+1}$, and the expansion begins at order $\alpha$. Note that Eq.~(\ref{2}) implies non-trivial relations among the coefficients $k_{nm}$ in Eq.~(\ref{1}). At order $n$, there are $2n+1$ coefficients $k_{nm}$, $0 \le m \le 2n$ in Eq.~(\ref{1}), but only $n+2$ coefficients $\tilde k_{nm}$, $0 \le m \le n+1$ in Eq.~(\ref{2}).

The right-hand-side (rhs) of Eq.~(\ref{2}) can be written in terms of the LL series $\rgel f_0(\alpha \rgel)=\tilde k_{12} \alpha \rgel^2+
\tilde k_{23} \alpha^2 \rgel^3 + \ldots$, the NLL series $f_1(\alpha \rgel)=\tilde k_{11} \alpha \rgel +
\tilde k_{22} \alpha^2 \rgel^2 + \ldots$, the NNLL series $\alpha f_2(\alpha \rgel)=\tilde k_{10} \alpha  +
\tilde k_{21} \alpha^2 \rgel + \ldots$ etc.\ as
\begin{eqnarray}
\log F_E &=&\rgel f_{0}(\alpha \rgel)+f_{1}(\alpha \rgel)+\alpha f_{2}(\alpha \rgel)
+ \ldots\, .
\label{3}
\end{eqnarray}
$f_0$ and $f_1$ begin at order $\alpha$, and the remaining $f_n$ begin at order one.

In this paper, LL, NLL, etc.\ (with no subscripts) will refer to the counting for $\log F_E$. This is also the counting appropriate for a renormalization group improved computation, and is different from the conventional counting discussed above. If one looks at the order $\alpha^2$ terms, for example, the conventional counting is that the $\rgel^4$ term is $\text{LL}_{\text{F}}$, the $\rgel^3$ term is $\text{NLL}_{\text{F}}$, the $\rgel^2$ term is $\text{N}^2\text{LL}_{\text{F}}$, the $\rgel$ term is $\text{N}^3\text{LL}_{\text{F}}$, and the $\rgel^0$ term is $\text{N}^4\text{LL}_{\text{F}}$. Using our counting, the terms are given by exponentiating $\log F_E$ to $\text{LL}$,  $\text{NLL}$, $\text{N}^2\text{LL}$, $\text{N}^2\text{LL}$, and $\text{N}^3\text{LL}$, respectively. At higher orders, the mismatch in powers of N between the two counting methods increases.

For precision electroweak studies, the first few orders are sufficient. Typical loop corrections are suppressed by $\alpha/(4 \pi)$. There can be large coefficients in the perturbation expansion. For example, there are large coefficients in the cusp anomalous dimension (see Eqs.~(\ref{101},\ref{3loopgam})). In this paper we have computed corrections to the Sudakov form factor; for dijet production and processes involving four-particle operators, the anomalous dimensions are at least twice as large as for the Sudakov problem. For these reasons, we use the estimate $\alpha$ instead of $\alpha/(4\pi)$ for the size of loop corrections. For QCD, $\alpha \sim 0.1$, and for electroweak corrections, $\alpha \to \aem/\sin^2\theta_W \sim0.03$. $\log s/M_Z^2 \sim 8$ for $s \sim 4$~TeV, so $\alpha \rgel \sim 1$ for QCD and $\sim 0.2$ for electroweak corrections. The NLL series is of order ten percent for QCD corrections, and a few percent for electroweak corrections. The NNLL series is of order a percent for QCD, and sub-percent for electroweak corrections.

\subsection{Infrared Evolution Equations}

An expression for the Sudakov form-factor in the limit $M/Q \ll 1$ with onshell massless fermions, $p_2^2 = p_1^2 = 0$, obtained using the evolution equations is~\cite{mueller,collins,sen,js} 
\begin{eqnarray}
&&\log F_E(Q^2) = \log F_0(a(M)) \nn
&&+ \int_{M^2}^{Q^2}\frac{{\rm d}\mu^2}{\mu^2}  \left[\zeta(a(\mu))+\xi(a(M))+ \int_{M^2}^{\mu^2} \frac{{\rm d} \mu^{\prime\, 2}}{\mu^{\prime\, 2}}
\Gamma(a(\mu^\prime))\right]\nn
\label{4}
\end{eqnarray}
in terms of functions $F_0$, $\zeta$, $\xi$ and $\Gamma$ of the coupling constant, which have the expansions
\begin{eqnarray}
F_0(a) &=& 1 + F_0^{(1)} a +  F_0^{(2)} a^2 + \ldots\, , \nn 
\Gamma(a) &=& \Gamma^{(1)} a +  \Gamma^{(2)} a^2 + \ldots\, , \nn
\zeta(a) &=& \zeta^{(1)} a +  \zeta^{(2)} a^2 + \ldots\, , \nn
\xi(a) &=& \xi^{(1)} a +  \xi^{(2)} a^2 + \ldots \, .
\label{9}
\end{eqnarray}
The superscript gives the loop-order of the Feynman graphs which contribute. $\Gamma$ is known as the cusp anomalous dimension.

The gauge coupling constant $g$ satisfies the renormalization group evolution equation
\begin{eqnarray}
\mu \frac{ {\rm d} g}{{\rm d}\mu}&=& \beta_g(g) \nn
&=&-b_0 \frac{g^3}{16 \pi^2} 
-b_1 \frac{g^5}{(16 \pi^2)^2}-b_2 \frac{g^7}{(16 \pi^2)^3}+\ldots \, ,\nn
\label{8}
\end{eqnarray}
or equivalently,
\begin{eqnarray}
\mu \frac{ {\rm d} a}{{\rm d}\mu}&=& \beta_a(a)\nn
&=&-2 a \left( b_0 a + b_1 a^2 + b_2 a^3 +\ldots\right) \, .
\label{10}
\end{eqnarray}
The one-loop coefficient is
\begin{eqnarray}
b_0 &=& \frac{11}3 C_A - \frac43 T_F n_F -\frac13 T_F n_s
\end{eqnarray}
where $n_s$ is the number of complex scalars, and $2n_F$ is the number of fermion weak doublets, in the convention of Ref.~\cite{js}.

After combining Eqs.~(\ref{4}--\ref{10}) and expanding to order $a(M)^3$,  the form factor takes the following form:
\begin{eqnarray}
&&\log F_E(Q^2) =  \nn
&&\ \   \left[ F_0^{(1)} +\left(\zeta^{(1)}+\xi^{(1)} \right)\rgel+ \frac{1}{2} \Gamma^{(1)} \rgel^2 \right]a(M) \nn
&&\ \ + \left[-\frac{1}{2} \left(F_0^{(1)} \right)^2+F_0^{(2)}+\left( \zeta^{(2)}+\xi^{(2)} \right)\rgel  \right. \nn
&& \ \  +\left. \frac{1}{2}\left( \Gamma^{(2)}-b_0\zeta^{(1)} \right)\rgel^2 -\frac{1}{6} \left( b_0\Gamma^{(1)} \right) \rgel^3 \right] a(M)^2\nn
&&\ \ + \left[\frac{1}{3} \left( F_0^{(1)} \right)^3 -F_0^{(1)}F_0^{(2)}+F_0^{(3)} + \left( \zeta^{(3)}+\xi^{(3)} \right)\rgel  \right. \nn
&&\ \ + \left. \frac{1}{2}\left( \Gamma^{(3)}-b_1\zeta^{(1)}-2b_0\zeta^{(2)} \right) \rgel^2 +\frac{1}{6} \left(  -b_1\Gamma^{(1)} \right. \right. \nn
&& \ \  +\left. \left.   2 b_0 \left( b_0 \zeta^{(1)}-\Gamma^{(2)}\right)\right) \rgel^3 + \frac{1}{12}b_0^2\Gamma^{(1)} \rgel^4 \right] a(M)^3\nn
&&+\ldots\, .
\end{eqnarray}
A comparison of this expansion with Eq.~(\ref{3}) shows that $f_0$ is determined by $\Gamma^{(1)}$ and  $b_0$, $f_1$ by $\Gamma^{(1,2)}, \zeta^{(1)}, \xi^{(1)}, b_{0,1}$, and 
$f_2$ by $F_0^{(1)}, \Gamma^{(1,2,3)}, \zeta^{(1,2)}, \xi^{(2)}, b_{0,1,2}$.  In general $f_n$ is determined by $\xi^{(n)}$ and  terms up to $F_0^{(n-1)}, \Gamma^{(n+1)}, \zeta^{(n)}, b_{n}$.

The expression Eq.~(\ref{4}) is not unique. The identity

\begin{eqnarray}
\hspace{-0.8cm}\frac12\int_{y^2}^{z^2} \frac{{\rm d}\mu^2}{\mu^2} \frac{\partial G(a(\mu))}{\partial a(\mu)} \beta_a(a(\mu)) &=& 
G(a(z))-G(a(y)) 
\label{6}
\end{eqnarray}
can be used to show that Eq.~(\ref{4}) is invariant under the transformation

\begin{eqnarray}
\Gamma(a(\mu')) &\to& \Gamma(a(\mu')) + \frac{\partial G(a(\mu'))}{\partial a} \beta_a(a(\mu'))\, ,\nn
\zeta(a(\mu)) &\to& \zeta(a(\mu))-2 G(a(\mu))\, ,\nn
\xi(a(M)) &\to& \xi (a(M)) +2G(a(M))\, .
\label{12}
\end{eqnarray}
As a result, $\Gamma$, $\zeta$ and $\xi$  are not uniquely determined from $F_E(Q^2)$ by Eq.~(\ref{4}).

\subsection{Renormalization Group Evolution Equations}

The corresponding expression for $F_E(Q^2)$ in the EFT formalism, as will be derived in Sec.~\ref{sec:massless}, is
\begin{eqnarray}
&&\log F_E(Q^2) = C(a(Q))+ D_0(a(M)) \nn
&&\qquad +  D_1 (a(M)) \log\frac{Q^2}{M^2} \nn
&&- \int_{M}^{Q}\frac{{\rm d}\mu}{\mu}  \left[A(a(\mu)) \log\frac{\mu^2}{Q^2}+ B (a(\mu))
\right]\, ,
\label{13}
\end{eqnarray}
where $\exp C(a(Q))$ is the multiplicative matching coefficient at $Q^2$, $\gamma(\mu)=A(a(\mu)) \log(\mu^2/Q^2)+B(a(\mu))$ is the SCET anomalous dimension between $Q$ and $M$, and $\exp D(a(M))$, $D(a(M))=D_0(a(M)) +  D_1 (a(M)) \log{Q^2}/{M^2}$ is the multiplicative matching coefficient at $M$. The matching coefficient $C$ and the SCET anomalous dimension $\gamma$ are independent of physics at the low scale $M$, and so do not depend on the gauge boson and Higgs masses. The new feature of the massive gauge boson calculation is the existence of a single-log term, $D_1(a(M))$ in the matching at $M$. That there are no higher powers of $\log Q^2/M^2$ in the matching is proved to hold to all orders in Sec.~\ref{sec:linear}. $A,B,C,D_{0,1}$ have loop expansions analogous to Eq.~(\ref{9}). The $\text{N}^n\text{LL}$ series for $\log F_E$ requires $A^{(n+1)}$, $B^{(n)}$, $D^{(n)}$, and $C^{(n-1)}$. $D^{(n)}$ contributes \emph{only} to the $\alpha^n \rgel$ term in $f_n$.

The identity Eq.~(\ref{6}) and
\begin{eqnarray}
 \int_{M^2}^{Q^2}\frac{{\rm d}\mu^2}{\mu^2}   \int_{M^2}^{\mu^2} \frac{{\rm d}\mu^{\prime\, 2}}{\mu^{\prime \, 2}}
\Gamma(a(\mu^\prime)) &=&  \int_{M^2}^{Q^2}\frac{{\rm d}\mu^2}{\mu^2}
\Gamma(a(\mu)) \log \frac{Q^2}{\mu^2}\nn
\label{14}
\end{eqnarray}
can be used to show that Eq.~(\ref{13}) is unchanged by the redefinitions
\begin{eqnarray}
A(a(\mu)) &\to& A(a(\mu)) + \frac{\partial \tilde G(a(\mu))}{\partial a} \beta_a(a(\mu))\, ,\nn
B(a(\mu)) &\to& B(a(\mu))+ \frac{\partial \tilde H(a(\mu))}{\partial a} \beta_a(a(\mu)) \nn && +2 \tilde G(a(\mu))\, , \nn
C(a(Q)) &\to& C(a(Q)) + \tilde H (a(Q))\, , \nn
D_0(a(M)) &\to& D_0(a(M)) - \tilde H (a(M))\, , \nn
D_1(a(M)) &\to& D_1(a(M))+ \tilde G(a(M))\, .
\label{16}
\end{eqnarray}
Transformations such as these can arise from a change of scheme in the computation of the SCET matching coefficients and anomalous dimensions.

We can now demonstrate the equivalence of the Sudakov form-factor in Eq.~(\ref{13}) and the form-factor given in Eq.~(\ref{4}).  By taking $G(a) = -\xi(a)/2$ in  Eq.~(\ref{12}),  $\tilde{H}(a) = -C(a)$ and $\tilde{G}(a) = -D_1(a)$ in Eq.~(\ref{16}),  brings  Eq.~(\ref{4}) with Eq.~(\ref{13}) to a common form, and gives the identifications:
\begin{eqnarray}
&&\frac12 A(a(\mu))-\frac12 \frac{\partial D_1(a(\mu))}{\partial a} \beta_a(a(\mu)) \nn
&&\qquad \qquad = \Gamma(a(\mu)) -\frac12\frac{\partial \xi(a(\mu))}{\partial a} \beta_a(a(\mu))\, , \nn
&&-\frac12 B(a(\mu)) +\frac12\frac{\partial C(a(\mu))}{\partial a} \beta_a(a(\mu)) +D_1(a(\mu)\nn \nn
&&\qquad \qquad= \zeta(a(\mu)) + \xi(a(\mu))\, ,\nn \nn
&&C(a(M))+ D_0(a(M)) = \log F_0(a(M)).
\label{7}
\end{eqnarray}
The lhs of Eq.~(\ref{7}) is invariant under Eq.~(\ref{16}), and the rhs under Eq.~(\ref{12}).
The computations of the SCET anomalous dimension and the cusp anomalous dimension in the literature use the same scheme, so that $\Gamma=A/2$, and
\begin{eqnarray}
\frac12 A(a)&=& \Gamma(a)\, ,\nn
D_1 (a)  &=& \xi(a) \, ,\nn
-\frac12 B(a) +\frac12\frac{\partial C(a)}{\partial a} \beta_a(a)  &=& \zeta(a)\, , \nn
C(a)+ D_0(a) &=& \log F_0(a)\, .
\label{15}
\end{eqnarray}

The expansion of $\log F_E(Q^2)$ to order $a(M)^3$ using the SCET form is:
\begin{eqnarray}
&&\log F_E(Q^2) =  \nn
&&\ \   \left[ C^{(1)} + D_0^{(1)} +\left(-\frac{1}{2}B^{(1)}+D_1^{(1)} \right)\rgel+ \frac{1}{4} A^{(1)} \rgel^2 \right]a(M) \nn
&&\ \ + \left[C^{(2)} + D_0^{(2)}+\left( -\frac{1}{2}B^{(2)}-b_0C^{(1)}+D_1^{(2)} \right)\rgel  \right. \nn
&& \ \  +\left. \frac{1}{4}\left( A^{(2)}+b_0B^{(1)} \right)\rgel^2 -\frac{1}{12} \left( b_0A^{(1)} \right) \rgel^3 \right] a(M)^2\nn
&&\ \ + \left[C^{(3)} + D_0^{(3)} + \left( -\frac{1}{2}B^{(3)}-b_1C^{(1)}-2b_0C^{(2)} \right. \right.  \nn
&&\ \  \left. \left. +D_1^{(3)} \right)\rgel  +\frac{1}{4}\left( A^{(3)}+b_1B^{(1)} \right. \right. \nn 
&& \left. +2b_0\left(B^{(2)}+2b_0C^{(1)}\right)  \right) \rgel^2 -\frac{1}{12} \left(  b_1A^{(1)} \right.  \nn
&& \ \  \left. \left. + 2b_0^2 B^{(1)}+2b_0 A^{(2)} \right) \rgel^3 + \frac{1}{24}b_0^2A^{(1)} \rgel^4 \right] a(M)^3\, .
\label{69}
\end{eqnarray}

\section{Massless External Particles}
\label{sec:massless}

In this section we calculate the form-factor $\log F_E(Q^2)$ for the case $Q^2 \gg M^2 \gg m_1^2, m_2^2$.  At scales $\mu > Q$ we use the full theory, and the renormalization group evolution of $c(\mu)$ is given by
\begin{eqnarray}
\mu \frac{ {\rm d} c(\mu)}{{\rm d}\mu}&=& \gamma_F(a(\mu))\ c(\mu)\, ,
\label{18}
\end{eqnarray}
where $\gamma_F(a) = \gamma_F^{(1)} a + \gamma_F^{(2)} a^2 +\ldots$ is the full theory anomalous dimension for $\mathcal{O}$.  The one-loop values $\gamma_F^{(1)}$ are given in Table~\ref{tab:results}.  The general form for $F_E$ given in Sec.~\ref{sec:expon} is for the vector current, where $\gamma_F=0$, and $c(\mu > Q )$ is chosen to be unity.
It also holds for the other operators with $c(\mu=Q)=1$ in the full theory.
\begin{table*}
\begin{eqnarray*}
\begin{array}{|c|r|c|c|c|}
\hline
\mathcal{O} & \gamma_F^{(1)}/C_F & C^{(1)}(\mu)/C_F & \gamma_1^{(1)}(\mu)/C_F & D^{(1)}(\mu)/C_F\\
\hline
\bar \psi  \psi & -6\quad &  - \lQ^2 +\frac{\pi^2}{6}-2&4 \lQ- 6 & -\lM^2+2\lM\lQ-3\lM+\frac92-\frac{5\pi^2}{6} \\[5pt]
\bar \psi \gamma^\mu \psi & 0\quad & -\lQ^2 +3\lQ+\frac{\pi^2}{6}-8&  4 \lQ- 6& -\lM^2+2\lM\lQ-3\lM+\frac92-\frac{5\pi^2}{6} \\[5pt]
\bar \psi \sigma^{\mu\nu} \psi & 2\quad & - \lQ^2+4\lQ+\frac{\pi^2}{6}-8 & 4 \lQ- 6& -\lM^2+2\lM\lQ-3\lM+\frac92-\frac{5\pi^2}{6} \\[5pt]
\phi^\dagger \phi & -6\quad & - \lQ^2+\lQ+\frac{\pi^2}{6}-2 &4 \lQ- 8 & -\lM^2+2\lM\lQ-4\lM+\frac{7}{2}-\frac{5\pi^2}{6} \\[5pt]
i(\phi^\dagger D^\mu \phi-D^\mu\phi^\dagger  \phi) & 0\quad & - \lQ^2 +4\lQ+\frac{\pi^2}{6}-8&  4 \lQ- 8 & -\lM^2+2\lM\lQ-4\lM+\frac{7}{2}-\frac{5\pi^2}{6} \\[5pt]
\bar \psi \phi, \phi^\dagger\psi & -3\quad &- \lQ^2+2 \lQ+\frac{\pi^2}{6}-4&  4 \lQ - 7&-\lM^2+2\lM\lQ-\frac72\lM+4-\frac{5\pi^2}{6} \\[5pt]
\hline
\end{array}
\end{eqnarray*}
\caption{\label{tab:results} One-loop corrections to the Sudakov form-factor. 
$\gamma_F$ is the full theory anomalous dimension, $C$ is the matching coefficient at $\mu \sim Q$, $\gamma_1$ is the SCET anomalous dimension, and $D$ is the matching coefficient at $\mu\sim M$. $\gamma^{(1)}_F$, $C^{(1)}$, $\gamma^{(1)}_1$ and $D^{(1)}$ are the coefficients of $a\equiv\alpha/(4\pi)$ in the one-loop corrections, and $\lQ\equiv\log Q^2/\mu^2$, $\lM\equiv\log M^2/\mu^2$.}
\end{table*}

The full theory is matched onto SCET at a scale $\mu$ of order $Q$.  The effective theory has modes with off-shellness of order $Q$ integrated out, so the matching coefficient depends on $\log Q^2/\mu^2$, and these logarithms are not large if $\mu \sim Q$. 

The operator $\mathcal{O}$ in the full theory matches to the operator $\widetilde\mathcal{O}$ in SCET:
\begin{eqnarray}
\bar \psi \Gamma \psi &\to& \exp C(\mu)\, [\bar \xi_{n,p_2} W_n] \Gamma
[W^\dagger_{\bar n} \xi_{\bar n,p_1}]\, , \nn
\phi^\dagger \phi &\to& \exp C(\mu)\, [ \Phi_{n,p_2}^\dagger W_n] 
[W^\dagger_{\bar n} \Phi_{\bar n,p_1}] \, ,\nn
i \phi^\dagger \darr{D}^\mu \phi &\to& \exp C(\mu)\, [ \Phi_{n,p_2}^\dagger W_n] (i\mathcal{D}_1+i\mathcal{D}_2)^\mu[W^\dagger_{\bar n} \Phi_{\bar n,p_1}] \, ,\nn
\bar \psi \phi &\to& \exp C(\mu)\, [\bar \xi_{n,p_2} W_n] 
[W^\dagger_{\bar n} \Phi_{\bar n,p_1}]\, ,
\label{19}
\end{eqnarray}
where $i\mathcal{D}_1 = \mathcal{P}+g( n \cdot A_{\bar n, q}) \frac{\bar n}{2}$,  $i\mathcal{D}_2 = \mathcal{P}^\dagger+g(\bar n \cdot A_{n, -q}) \frac{n}{2}$ and $\mathcal{P}$ are the SCET label operators introduced in Bauer {\it et al.\ }~\cite{SCET}.  
Collinear gauge invariance requires that, in the matching of gauge invariant operators at leading order in the power counting, the fields occur in the combination $W_n^\dagger \xi_{n,p}$, $W_n^\dagger \Phi_{n,p}$, where $W_n$ is a Wilson line containing $n$-collinear gauge fields obtained by integrating over a path in the $\bar n$-direction~\cite{SCET}.

\begin{figure}
\begin{center}
\includegraphics[width=4cm]{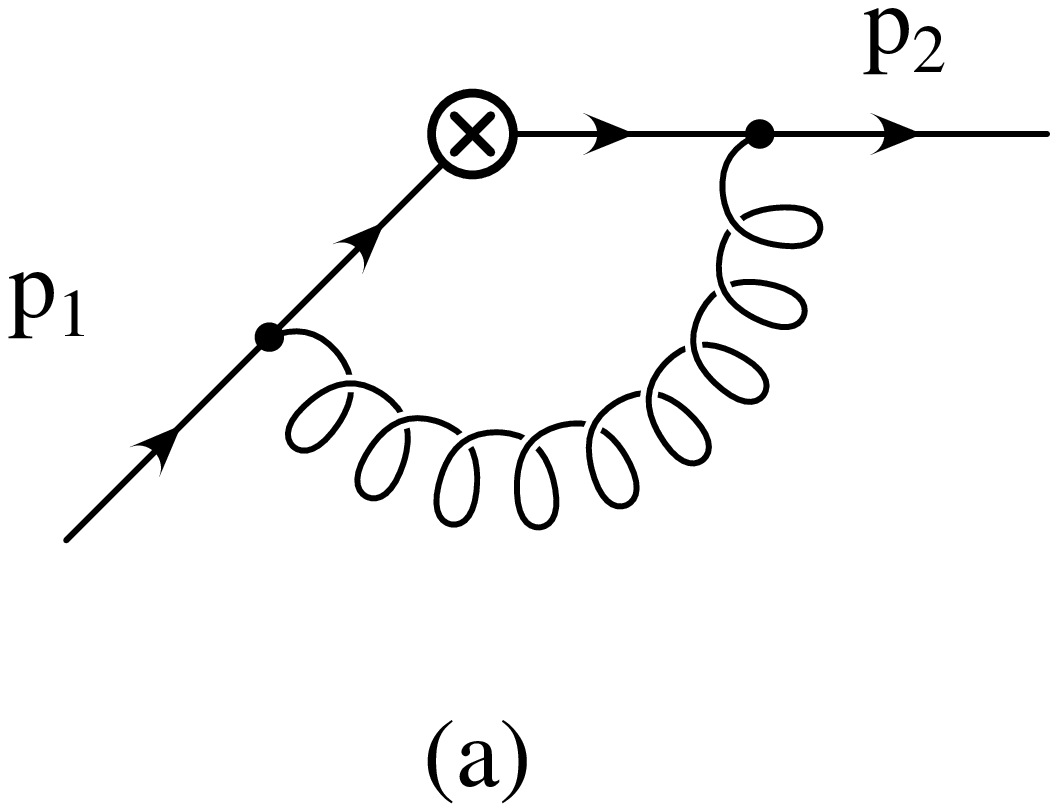}\includegraphics[width=4cm]{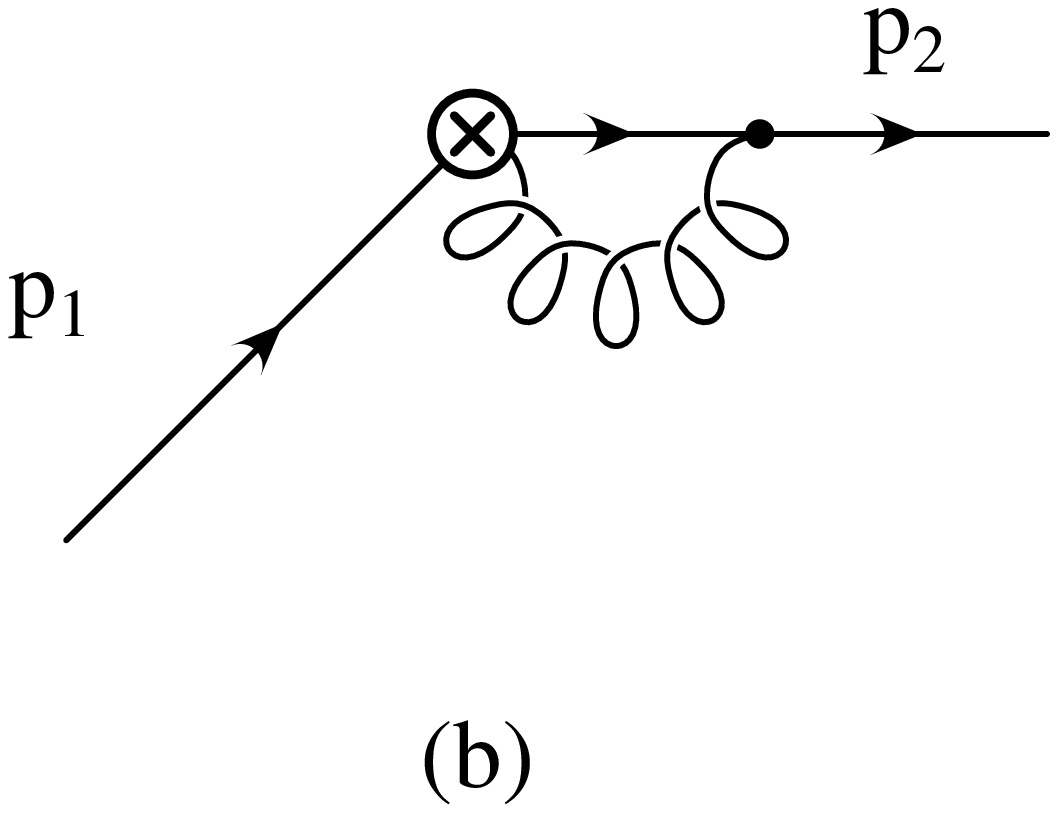}
\end{center}
\caption{\label{fig:match} Graphs contributing to the matching condition $C(\alpha(Q))$.
The solid line can be either a fermion or scalar.
The second graph only exists for the scalar case $\mathcal{O}=i(\phi^\dagger D^\mu \phi-D^\mu\phi^\dagger  \phi)$. }
\end{figure}

\begin{figure}
\begin{center}
\includegraphics[width=4cm]{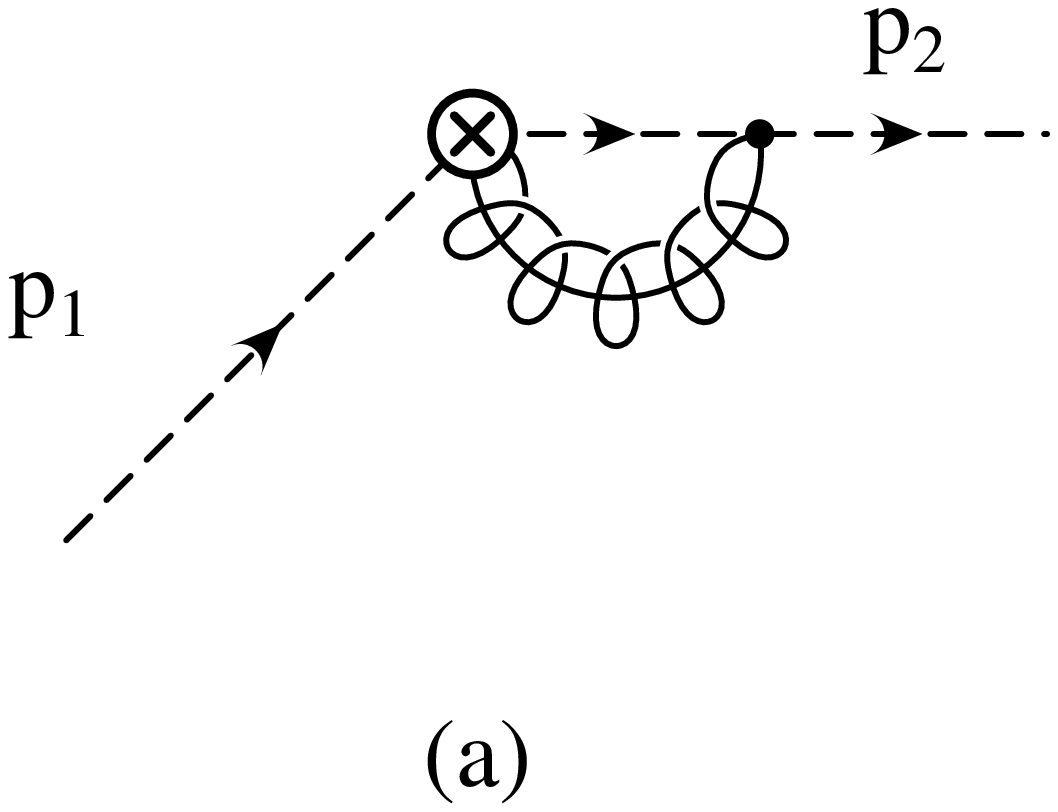}\includegraphics[width=4cm]{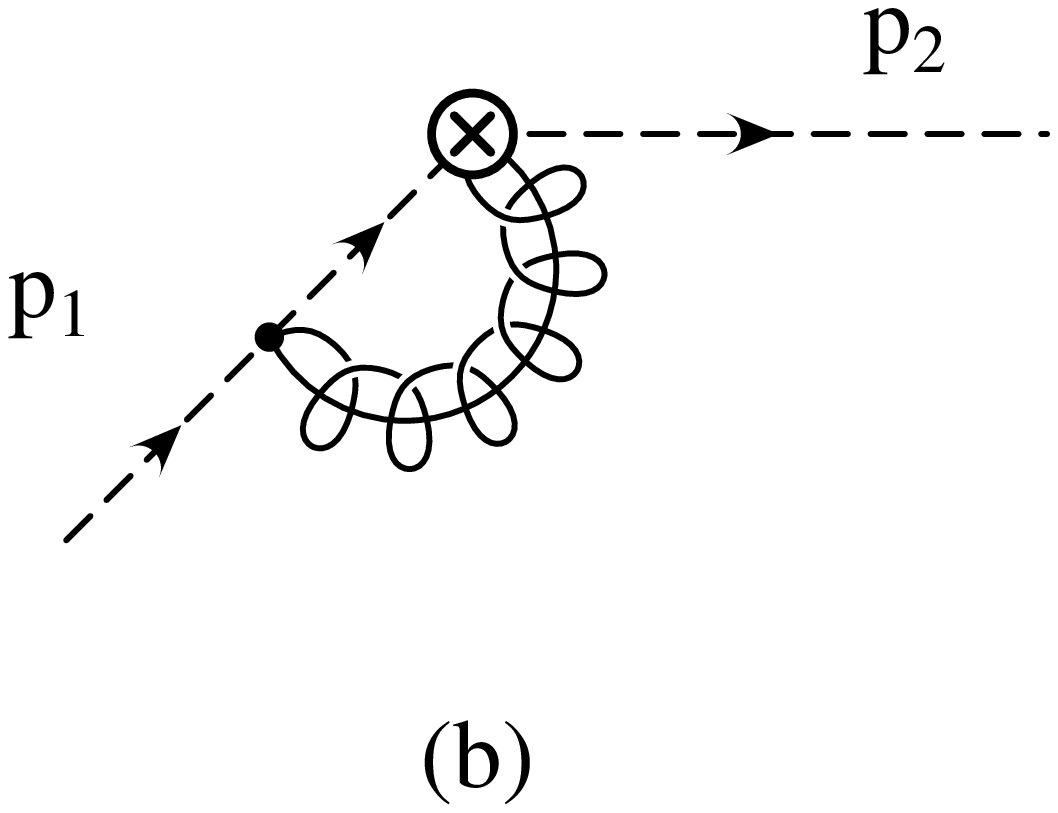}
\qquad\includegraphics[width=4cm]{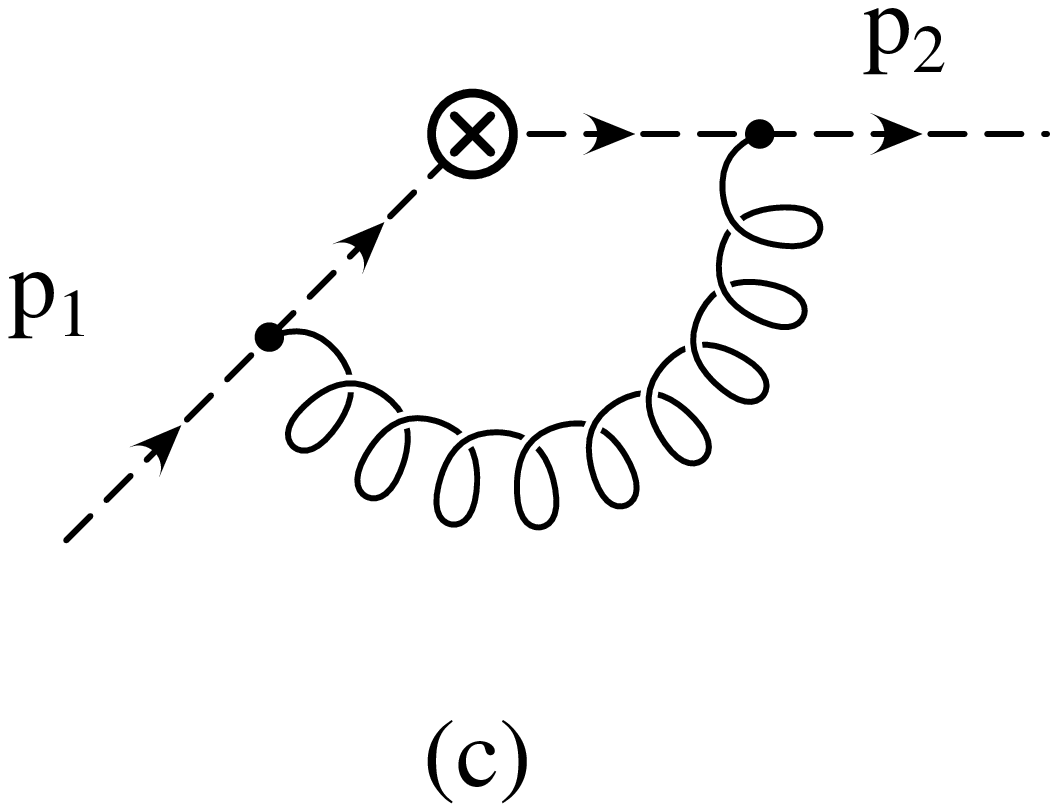}
\end{center}
\caption{\label{fig:scet} SCET graphs for the matrix element of $\widetilde\mathcal{O}$. The dotted lines are SCET propagators, and represent either fermions or scalars. 
The upper graphs are the $n$-collinear and $\bar n$-collinear graphs, and the lower graph is the ultrasoft graph. There are also wavefunction graphs. For $i\phi^\dagger \darr{D}^\mu \phi$, graph (a) also has a contribution where the gauge boson field at $\otimes$ arises from the covariant derivative. }
\end{figure}

$C(\mu)$ depends on the operator being matched (i.e.\ the $C$'s in Eq.~(\ref{19}) have different values, and $C$ can depend on $\Gamma$) and, for convenience, we have written the multiplicative matching coefficient as $\exp C(\mu)$ rather than $C(\mu)$.  As is well-known, the matching coefficient can be computed as the finite part of the full theory matrix element, evaluated on-shell, with all infrared scales such as the gauge boson mass set to zero (see e.g.~\cite{hqet,eft,dis}).  The full theory graphs to be evaluated at one-loop are those in Fig.~\ref{fig:match}, 
and when combined with the wavefunction and tree-level graphs, give the value of the full theory matrix element $\braket{p_2|\mathcal{O}|p_1}$.  The graphs for the EFT vertex correction are shown in Fig.~\ref{fig:scet}, and when combined  with the tree-level and wavefunction graphs, give the EFT matrix element $\braket{p_2|\widetilde{\mathcal{O}}|p_1}$.  The gauge boson and fermion masses are infrared scales and can be set to zero in the matching computation thus leading to scaleless integrals for the one-loop EFT and wavefunction graphs.  Since these scaleless integrals are set to zero in dimensional regularization, the EFT matrix element is equal to its tree-level value. The full theory and EFT operators $\mathcal{O}$ and $\widetilde\mathcal{O}$ are normalized to have the same tree-level value, so $\exp[C(\mu)] = \braket{p_2|\mathcal{O}|p_1}/\braket{p_2|\mathcal{O}|p_1}_{\text{tree}}$, i.e.\ the matching condition $\exp C$ is given by the on-shell full theory matrix element normalized to its tree-level value (see e.g. Ref.~\cite{hqet} for more details).

The computation of the SCET one loop graphs for $\mathcal{O}=\bar\psi \gamma^\mu \psi$ is identical to that for DIS~\cite{dis}.  Particle masses, such as the gauge boson mass, are all much smaller than $Q$, and only contribute $M^2/Q^2$ power corrections at the scale $Q$, which are being neglected.
The one-loop values of $C(\mu)$ for the other cases are computed similarly, and are given in Table~\ref{tab:results}, where $C(\mu)=C^{(1)}\alpha(\mu)/(4\pi)$ defines the one-loop correction $C^{(1)}$. There are no large logarithms in this matching correction if the matching scale $\mu$ is chosen to be of order $Q$.  We will choose the matching at the high scale to be at $\mu=Q$, and $C(\mu=Q)$ is given by the third column in Table~\ref{tab:results} with $\lQ\to0$.

The renormalization group evolution of $c(\mu)$ in the effective theory
is given by the anomalous dimension of $\widetilde\mathcal{O}$ in SCET. The anomalous dimension $\gamma_1$ is used to evolve $c(\mu)$ from $\mu=Q$ down to the low scale $\mu=M$. The one-loop anomalous dimension is given by the ultraviolet counterterms for the SCET graphs in Fig.~\ref{fig:scet} (after zero-bin subtraction, see Ref~\cite{zerobin}). 
As noted earlier, anomalous dimensions in SCET can depend on $Q$. Ultraviolet divergences do not depend on the infrared properties of the theory, such as a gauge boson mass, so the anomalous dimension for $\mathcal{O}=\bar \psi \gamma^\mu\psi $ is identical to the DIS result~\cite{dis}.  The same argument as that given in Ref.~\cite{dis} for deep inelastic scattering shows that $\gamma_1(\mu)$ is linear in $\log \mu^2/Q^2$ to all orders~\cite{dis,Bauer:2003pi}, so $\gamma$ is written as
\begin{eqnarray}
\gamma_1(\mu) &=& A(\alpha(\mu)) \log \frac{\mu^2}{Q^2} + B(\alpha(\mu))\, ,
\label{20}
\end{eqnarray}
which defines $A$ and $B$. The anomalous dimension has the expansion $\gamma_1=\gamma^{(1)}_1 a + \gamma^{(2)}_1 a^2 + \ldots$,   $A=A^{(1)} a + A^{(2)} a^2 + \ldots$, $B=B^{(1)} a + B^{(2)} a^2 + \ldots$
 The computations for the other cases are similar, and the results are given in Table~\ref{tab:results}. Note that the anomalous dimension depends only on the external fields for the operators, and is equal for the three fermion operators, and for the two scalar operators. The reason is that the EFT anomalous dimension depends on the IR divergence of the full theory graph, and the IR divergence is independent of the vertex factors. The anomalous dimension for $\bar \psi \phi$ is the average of the anomalous dimensions for the fermionic and scalar operators.

The next step in the EFT computation is the matching condition at the low scale $\mu \sim M$. At this scale, the massive gauge boson is integrated out, and one matches to an effective theory which is SCET without the massive gauge boson. In our toy example, this effective theory contains no gauge particles, and is a free theory. There is no need to introduce any propagating gauge modes below $M$~\cite{ir}.  The matching at $\mu \sim M$ is given by evaluating the graphs in Fig.~\ref{fig:scet}, and the wavefunction graphs. The gauge boson mass can no longer be set to zero, since it is of the same order as the matching scale, and the one loop SCET graphs are non-zero. The matching computation is discussed in detail here for the fermion vector current. The other cases are treated similarly. 

One matches the operator $c(\mu)[\bar \xi_{n,p_2} W_n] \gamma^\mu [W^\dagger_{\bar n} \xi_{\bar n,p_1}]$ in SCET with gauge particles (the theory above $M$) onto the operator $\left[\exp D(\mu)\right]c(\mu) \bar\xi_{n,p_2} \gamma^\mu\xi_{\bar n, p_1}$ in SCET without gauge particles (the theory below $M$).  The $n$-collinear graph in Fig.~\ref{fig:scet} gives
\begin{eqnarray}
I_n &=&-i g^2 \mu^{2\epsilon} C_F c(\mu) \int { {\rm d}^d k \over (2 \pi )^d} \frac{1}{k^2-M^2}\nn
&\times&  {\slashed{\bar n}\  \over 2}  n^\alpha \frac{\slashed{n} }{2} \frac{\bar n \cdot (p_2-k) }{(p_2 - k )^2} \gamma^\mu \frac{1}{- \bar n \cdot k}\bar n_\alpha  \nn
&=& -2 i g^2 C_F \gamma^\mu \mu^{2\epsilon} \int { {\rm d}^d k \over (2 \pi )^d} \nn
&\times&  
   \frac{\bar n \cdot (p_2-k)}{[(p_2 - k )^2+i0^+][- \bar n \cdot k+i0^+][k^2-M^2+i0^+]}. \nn
\label{21}
\end{eqnarray} 
This integral is divergent, even in $4-2\epsilon$ dimensions with an off-shellness, unlike the previously studied examples where the gauge boson was massless. A related divergence was encountered by Beneke and Feldman in their study of the $B \to \pi \ell \nu$ form-factor. Beneke and Feldman used an analytic regulator~\cite{bf,analytic}  to evaluate their integrals, and we use an extension of their method. A similar procedure was used by Jantzen {\it et al.\ }\cite{jkps4} in their study of two-loop electroweak Sudakov corrections. The $p_i$ propagator denominator $(p_i-k)^2$ in the full theory  is analytically continued to
\begin{eqnarray}
\frac{1}{(p_i-k)^2} &\to& \frac{ (-\nu_i^2)^{\delta_i}}
{\left[(p_i-k)^2\right]^{1+\delta_i}}\: .
\label{22}
\end{eqnarray}
where $\nu_i$ and $\delta_i$ are new parameters. The $(p_2-k)^2$ denominator in Eq.~(\ref{21}) arises from the collinear $p_2$ propagator, and so gets modified as in Eq.~(\ref{22}). The $-\bar n \cdot k$ propagator in Eq.~(\ref{21}) arises from the $(p_1-k)^2$ propagator when $k$ becomes $n$-collinear. In this limit
\begin{eqnarray}
\frac{1}{(p_1-k)^2} &\to& \frac{ (-\nu_1^2)^{\delta_1}}
{\left[(n \cdot p_1) (-\bar n \cdot k)\right]^{1+\delta_1}}\: .
\label{23}
\end{eqnarray}
We will therefore analytically continue the $-\bar n \cdot k$ propagator in Eq.~(\ref{21}), which arises from the $W_n$ Wilson line in $\mathcal{O}$ using
\begin{eqnarray}
\frac{1}{-\bar n \cdot k} &\to& \frac{(-\nu_1^-)^{\delta_1}}
{(-\bar n \cdot k)^{1+\delta_1}}\, ,
\label{24}
\end{eqnarray}
where $\nu_1^- \equiv \nu_1^2/ p_1^+$. We will see below that it is important that $\nu_1^-$ is related to $\nu_1^2$ in this way. Note that under boosts, $\nu_1^-$ transforms like the minus component of a four-vector. With this choice, Eq.~(\ref{21}) gives
\begin{eqnarray}
I_n   &=&  -2{\alpha \over 4 \pi}C_F c(\mu) \gamma^\mu \left(\frac{\mu^2}{M^2}\right)^{\epsilon} \left(\frac{\nu_2^2}{M^2}\right)^{\delta_2}  \left( \frac{\nu_1^-}{p_2^-}\right)^{\delta_1}      \nn
&& \times \frac{\Gamma(\epsilon+\delta_2)}{\Gamma(1+\delta_2)} \frac{\Gamma(2-\epsilon-\delta_2)\Gamma(\delta_2-\delta_1)}{\Gamma(2-\epsilon-\delta_1)}\ .
\label{25}
\end{eqnarray}
The regulated value of $I_n$ is given by setting $\delta_i = r_i \delta$ and taking the limit $\delta \to 0$ first, followed by $\epsilon \to 0$~\cite{bf,analytic},
\begin{eqnarray}
I_n&=&\frac{\alpha}{4\pi}C_F  c(\mu) \gamma^\mu \Biggl[ \frac{2}{r_1-r_2}\frac{1}{\delta \epsilon}
+\frac{2}{r_1-r_2} \frac{1}{\delta} \log \frac{\mu^2}{M^2}\nn
&&-\frac{2r_2}{r_1-r_2}\frac{1}{\epsilon^2}\nn
&& +\frac{1}{\epsilon}\biggl(2 + \frac{2r_1}{r_1-r_2} \log \frac{\nu_1^-}{p_2^-}
+\frac{2r_2}{r_1-r_2} \log \frac{\nu_2^2}{\mu^2} \biggr) \nn
 &&+2+2 \log \frac{\mu^2}{M^2}+\frac{2r_2}{r_1-r_2} \log \frac{\mu^2}{M^2}
 \log \frac{\nu_2^2}{\mu^2}\nn
 &&+\frac{2r_1}{r_1-r_2} \log \frac{\mu^2}{M^2}
 \log \frac{\nu_1^-}{p_2^-} + \frac{r_2}{r_1-r_2} \log^2 \frac{\mu^2}{M^2}\nn
&& +\frac{r_2 \pi^2}{2(r_1-r_2)}-\frac{r_1 \pi^2}{3(r_1-r_2)}
\Biggr],
\label{26}
\end{eqnarray}
which is a boost invariant expression, since $\nu_1^-/p_2^-$ is boost invariant. Equation~(\ref{26}) is valid away from the symmetric point $r_1=r_2$.

The $\bar n$-collinear graph is given by Eq.~(\ref{26}) with the replacements $\delta_1 \leftrightarrow \delta_2$, $\nu_2 \to \nu_1$, $\nu_1^- \to \nu_2^+$, $p_2^- \to p_1^+$, with $\nu_2^+ \equiv \nu_2^2/p_2^-$,
\begin{eqnarray}
I_{\bar n}   &=&  -2{\alpha \over 4 \pi}C_F c(\mu) \gamma^\mu \left(\frac{\mu^2}{M^2}\right)^{\epsilon} \left(\frac{\nu_1^2}{M^2}\right)^{\delta_1}  \left( \frac{\nu_2^+}{p_1^+}\right)^{\delta_2}      \nn
&& \times \frac{\Gamma(\epsilon+\delta_1)}{\Gamma(1+\delta_1)} \frac{\Gamma(2-\epsilon-\delta_1)\Gamma(\delta_1-\delta_2)}{\Gamma(2-\epsilon-\delta_2)}\ .
\label{25b}
\end{eqnarray}
The parameters $\nu_2^+$ and $\nu_1^-$ play the same role as $\mu^\pm$ in the rapidity regularization method of Ref.~\cite{zerobin}.

The ultrasoft graph in Fig.~\ref{fig:scet} is regulated by the same method. The $p_2$ propagator $(p_2-k)^2$ is multipole expanded in the effective theory, and becomes
$-p_2^- k^+$, where $p_2^-$ is a label momentum. Using Eq.~(\ref{22}) for the fermion propagators, we see that after multipole expansion, they are regulated in the same way as the Wilson line propagators. The ultrasoft graph gives
\begin{eqnarray}
I_{\rm us} &=&-ig^2C_Fc(\mu) \gamma^\mu \int { {\rm d}^d k \over (2 \pi )^d} {1 \over k^2-M^2} \nn
&&n^\alpha {(-\nu_2^+)^{\delta_2}  \over [n \cdot (p_2-k)]^{1+\delta_2}} \gamma^\mu
{ (-\nu_1^-)^{\delta_1} \over [\bar n \cdot (p_1 -k )]^{1+\delta_1}}
\bar n_\alpha\, ,\nn
\label{27}
\end{eqnarray} 
and vanishes on-shell, since $p_2^+=p_1^-=0$. 

\begin{figure}
\begin{center}
\includegraphics[width=4cm]{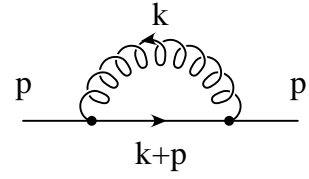}
\end{center}
\caption{\label{fig:wavef} One-loop correction to the fermion propagator.}
\end{figure}

\begin{figure}
\begin{center}
\includegraphics[width=3.5cm]{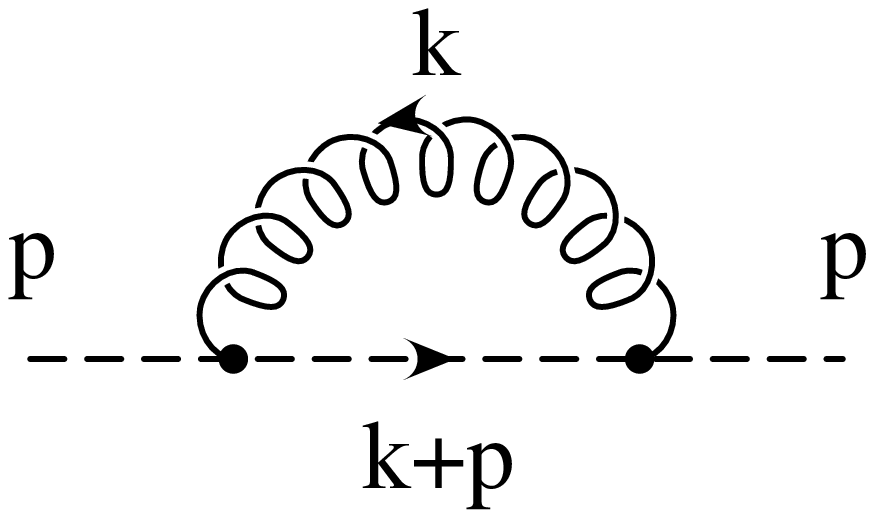}\qquad
\raise0.8cm\hbox{\includegraphics[width=3.5cm]{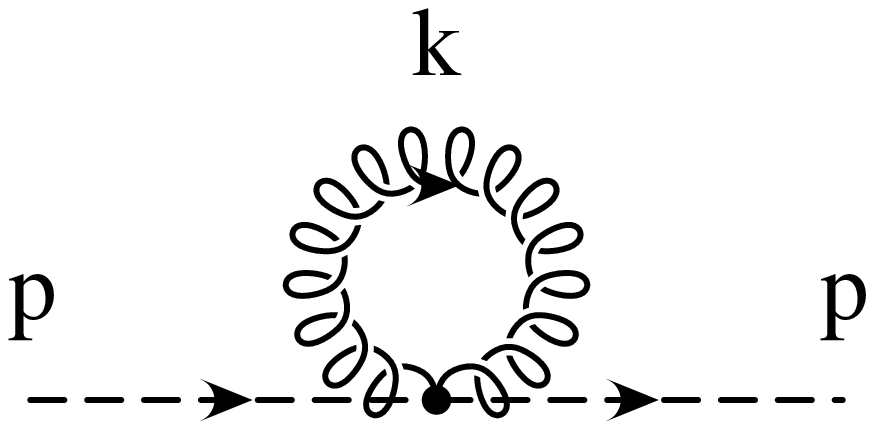}}
\end{center}
\caption{\label{fig:waves} One-loop correction to the scalar propagator.}
\end{figure}

The total SCET contribution is given by the sum of the $n$-collinear, $\bar n$-collinear and ultrasoft graphs, as well as the wavefunction renormalization correction. The collinear correction to the particle propagator is the same as in the full theory~\cite{SCET}, and the ultrasoft correction vanishes, so the wavefunction corrections are the same as in the full theory. The fermion graph, Fig.~\ref{fig:wavef}, gives
\begin{table}
\begin{eqnarray*}
\begin{array}{|c|c|c|c|c|}
\hline
\text{Field} & m & M & \\[5pt]
\hline
\psi & 0 & 0 & \frac{1}{\eUV}-\frac{1}{\eIR} \\[5pt]
\psi & 0 & \not=0 & \frac{1}{\eUV}-\frac12 -\lM \\[5pt]
\psi & \not=0 & 0 & \frac{1}{\eUV}+\frac{2}{\eIR}+4 -3 \lm \\[5pt]
\psi & \not=0 & \not=0 &  \frac{1}{\eUV}-\frac12 -\lM + h_F(m^2/M^2) \\[5pt]
\phi & 0 & 0 & -\frac{2}{\eUV}+ \frac{2}{\eIR} \\[5pt]
\phi & 0 & \not=0 & -\frac{2}{\eUV}-\frac32+ 2 \lM \\[5pt]
\phi & \not=0 & 0 & -\frac{2}{\eUV}+\frac{2}{\eIR} \\[5pt]
\phi & \not=0 & \not=0 &  -\frac{2}{\eUV}-\frac32+ 2 \lM + h_S(m^2/M^2)\\[5pt]
h_v &\infty& 0 & -\frac{2}{\eUV}+ \frac{2}{\eIR}\\[5pt]
h_v &\infty& \not=0 & -\frac{2}{\eUV}+ 2 \lM \\[5pt]
\hline
\end{array}
\end{eqnarray*}
\caption{\label{tab:wave} One-loop gauge boson contribution to on-shell wavefunction renormalization. The gauge boson mass is $M$ and the particle (fermion or scalar) mass is $m$. $h_{F,S}$ are given in Appendix~\ref{app:integrals}.}
\end{table}
\begin{eqnarray}
C_F {\alpha \over 4 \pi}  \, i \slashed{p}\left[ {1 \over \eUV} -\frac12 - \ln{M^2 \over  \mu^2 } \right]\, ,
\label{31}
\end{eqnarray}
and so contributes a wavefunction correction
\begin{eqnarray}
\delta z = C_F {\alpha \over 4 \pi}  \, \left[ {1 \over \eUV} -\frac12 - \lM  \right]\, .
\end{eqnarray}
Our normalization convention is such that the on-shell matrix element gets a contribution $-\delta z/2$ for each external particle. The wavefunction corrections for the various cases we need are tabulated in Table~\ref{tab:wave}. In the table, we have distinguished between UV and IR divergences by the subscript on the $1/\epsilon$ terms. The scalar operators requires the scalar propagator correction, Fig.~\ref{fig:waves}, which gives
\begin{eqnarray}
&&ip^2 \frac{\alpha_s}{4\pi} C_F  \Biggl[-\frac{2}{\eUV}
 -\frac32 + 2 \ln \frac{M^2}{\mu^2}\Biggr]\nn
&& -iM^2 \frac{\alpha_s}{4\pi} C_F   \Biggl[-\frac{3}{\eUV}
 -1 +  3 \ln \frac{M^2}{\mu^2}\Biggr]\, .
\label{32}
\end{eqnarray}
The first term gives the wavefunction correction, and the second is the mass-shift of the scalar proportional to the gauge boson mass. The scalar mass-shift is canceled by the bare mass term in the scalar Lagrangian, which is adjusted to keep the physical scalar massless. This cancellation is an example of fine-tuning required to have light scalars.

The total on-shell amplitude $I_n+I_{\bar n}+I_s-2(\delta z/2)$ is
\begin{eqnarray}
&&\frac{\alpha}{4\pi}C_F c(\mu) \gamma^\mu\Biggl[ \frac{2}{\epsilon^2} +\frac{1}{\epsilon}\left(3-2 \lQ\right)+2 \lM \lQ \nn
&&\hspace{2.25cm}  -\lM^2 -3 \lM+\frac92  -\frac{5 \pi^2}{6} \Biggr].
\label{28}
\end{eqnarray}
The total amplitude Eq.~(\ref{28}) is independent of $\delta$, $r_1$ and $r_2$ introduced by the analytic regulator, and depends only on $\epsilon$ of dimensional regularization. The cancellation of the $\nu$ and $\delta$ dependence is discussed in more detail in Appendix~\ref{app:analytic}.
In evaluating Eq.~(\ref{28}), we have used $\nu_1^-=\nu_1^2/p_1^+$, $\nu_2^+=\nu_2^2/p_2^-$ and $Q^2=p_1^+ p_2^-$.
The $1/\epsilon$ and $1/\epsilon^2$ poles are ultraviolet divergences, and are  canceled by the renormalization counterterms in the effective theory. The IR divergences in the EFT are regulated by the gauge boson mass, so the $1/\epsilon$ divergences in Eq.~(\ref{28}) are UV divergences. The $1/\epsilon$ term multiplied by $-2$ gives the SCET anomalous dimension listed in Table~\ref{tab:results}, and is a non-trivial check on the analytic regulator computation. The SCET anomalous dimension was computed in Ref.~\cite{dis} using an off-shell regulator, and the analytic regulator gives the same result. While the total anomalous dimension is the same, the contribution of individual diagrams to the anomalous dimension depends on the regulator. For example, in Ref.~\cite{dis}, the ultrasoft graph had a $1/\epsilon$ divergence which contributed to the anomalous dimension, whereas the ultrasoft graph vanishes on-shell when evaluated using the analytic regulator method. Contributions can be moved between the collinear and ultrasoft diagrams, depending on the choice of regulator.

The EFT below the matching scale $\mu \sim M$ is SCET without gauge particles; thus there are no one loop diagrams to consider in the theory below $M$.  The finite part of Eq.~(\ref{28}) gives the multiplicative matching coefficient $\exp D(\mu)$ at the low scale $\mu$ of order $M$. The coefficient of $\widetilde \mathcal{O}$ in the effective theory after integrating out the gauge bosons is given by $c(\mu-0^+)=\left[\exp D(\mu)\right] c(\mu+0^+)$. The coefficient $D(\mu)$ has the usual expansion $D=D^{(1)}a + D^{(2)}a^2 + \ldots$, and at one-loop order is
\begin{eqnarray}
D^{(1)}&=&C_F \Biggl[  -\lM^2 +2 \lM \lQ -3 \lM+\frac92  
-\frac{5 \pi^2}{6} \Biggr]\, .\nn
\label{33}
\end{eqnarray}
for the fermion vector current. The other cases are computed similarly, and are given in Table~\ref{tab:results}. The matching at $M$ is independent of the vertex structure, and depends only on whether the particles are fermions or scalars. The $\bar \psi \phi$ matching is the average of the results for two fermions and two scalars. This is a new feature of the effective theory, which follows because the graphs factorize into contributions from the individual particles. The matching at $Q$ from the full theory does not have this property.

Note that $D(\mu)$ is a function of both $\lM$ and $\lQ$, and is linear in $\lQ$. The matching condition depends on both scales $Q$ and $M$. The dependence of the matching on the high scale $Q$ is a new feature of SCET with massive gauge bosons, and has not occurred in previous computations in SCET, or other EFTs. We noted earlier that SCET graphs know about the scale $Q$ through the labels $\bar n \cdot p_2$ and $n \cdot p_1$ since $Q^2=\bar n \cdot p_2\, n \cdot p_1$. Nevertheless, in previous computations such as DIS, the matching condition at the jet scale $M_J^2 \ll Q^2$ depended on $\log M_J^2/\mu^2$, and there was no $\log Q^2/\mu^2$ dependence. It is easy to see why there must in general be $\lQ$ terms in the matching condition in our case. 
If $D$ is the matching condition at $\mu$, and $\gamma_{h,l}$ are the anomalous dimension in the theories above and below $\mu$,
\begin{eqnarray}
\mu \frac{ \rd D}{\rd \mu} = \gamma_l(\mu)-\gamma_h(\mu).
\end{eqnarray}
In our example, $\gamma_l=0$ since the theory below $M$ is a free theory. Since $\gamma_h$ has the form Eq.~(\ref{20}) with a $\lQ$ term, such terms must also be present in $D$. Let us contrast this with DIS. For moments $M_N$ of the deep inelastic scattering structure function with $N \gg 1$, the jet scale is $M_J^2=Q^2/N$. The theory above the jet scale has an anomalous dimension $\gamma_h$ which depends on $\lQ$, and the theory below the jet scale has Altarelli-Parisi evolution with anomalous dimension $\gamma_l$ which depends on $\log N$. The two anomalous dimension are related in such a way that $\gamma_l-\gamma_h \propto \log Q^2/N$, the logarithm of the jet scale. The matching $D$ also depends only on the jet scale $Q^2/N$, and there are no  large logarithms in $D$ if $\mu$ is chosen to be of order the jet scale~\cite{dis}.

The $\lQ$ term in Eq.~(\ref{33}) is multiplied by $\lM$, and so there is no $\lQ$ term in $D$ if the matching scale is chosen to be exactly equal to $M$. This is accidental, and does not happen at higher orders. One can show explicitly that at two-loops, there is a non-zero $\lQ$ contribution to $D$ even if $\mu=M$. In the standard model, it is convenient to integrate the weak gauge bosons out at a single scale $\mu=M_Z$, and one has one-loop terms in the matching condition of the form $(\log Q^2/M_Z^2)( \log M_W^2/M_Z^2)$.

Renormalization group improved perturbation theory is used to sum logarithms in an EFT. This would not be possible if there were arbitrary powers of $\lQ$ in the matching condition. We will prove in Sec.~\ref{sec:linear} that to all orders in perturbation theory, the matching condition $D$ is \emph{linear} in $\lQ$. Thus renormalization group summation can be used to obtain all logarithms except the first, so that in the Sudakov problem at order $\alpha^n$, the $2n-1$ terms $\alpha^n \rgel^{2n}, \ldots, \alpha^n \rgel^2$ can be obtained by renormalization group evolution, but not the single log term $\alpha^n \rgel$, which gets a matching contribution from $D$. The general form for $D(\mu)$ is
\begin{eqnarray}
D(\mu) &=& D_0(a(\mu),\lM)+D_1(a(\mu),\lM)\lQ\, ,
\label{34}
\end{eqnarray}
which defines $D_{0,1}$. At one-loop, Eq.~(\ref{33}) gives
\begin{eqnarray}
D^{(1)}_0&=&C_F \Biggl[  -\lM^2  -3 \lM+\frac92  
-\frac{5 \pi^2}{6} \Biggr]\, ,\nn
D^{(1)}_1&=& 2 C_F \lM\, .
\end{eqnarray}
Choosing $\mu=M$ gives the matching coefficient
\begin{eqnarray}
D(\mu=M) &=& D_0(a(M),0)+D_1(a(M),0)\log \frac{Q^2}{M^2}\, .\nn
\label{35}
\end{eqnarray}
In our example, $D_1(a(M),0)=0$, so there is no $\log Q^2/M^2$ term in the one-loop matching coefficient. One expects that $D_1(a(M),0)\not=0$ at higher orders, so there can be a single large logarithm in the matching coefficient.

The final step in the computation is to compute the on-shell matrix element of $\widetilde \mathcal{O}$ in the theory below $M$. Since the gauge bosons have been integrated out, this theory is a free theory, and the matrix element is trivial, being given by its free field value. The Sudakov form-factor is defined as the ratio of the scattering amplitude to its value in the free theory, so the low-energy matrix element contribution to the Sudakov form-factor is unity.

The contributions to the Sudkakov form-factor are:
\begin{enumerate}
\item The coefficient $c(\mu)$ in the full theory just above the matching scale $\mu=Q$, which is chosen to be unity.
\item The multiplicative matching coefficient $\exp C(\mu)$ for the matching between the full theory and SCET at the scale $\mu = Q$.
\item The integral of the SCET anomalous dimension between $\mu=Q$ and $\mu=M$.
\item The multiplicative matching coefficient $\exp D(\mu)$ for the matching  at the scale $\mu = Q$ between SCET, and SCET with the gauge bosons integrated out.
\item The low-energy matrix element, which gives unity, using the conventional normalization for the form-factor.
\end{enumerate}
Combining these contributions gives Eq.~(\ref{13}) for the Sudakov form-factor given earlier. The terms are represented schematically as:
\begin{eqnarray}
\begin{array}{ccccccc}
C & \gamma_1 & D \\
Q & \longrightarrow &  M 
\end{array}
\end{eqnarray}

The expression Eq.~(\ref{13}) for the Sudakov form-factor, with the one-loop coefficients given in Table~\ref{tab:results},  can be compared with known fixed order results in the case of the fermion vector current by expanding this in a power series expansion in $\alpha(M)$ as shown in Eq.~(\ref{69}). The result correctly reproduces the known $\alpha \rgel$, $\alpha^2 \rgel^4$ and $\alpha^2 \rgel^3$ terms ($\rgel=\log Q^2/M^2$).

Comparison with the two-loop results of Ref.~\cite{jkps4,js} allows us to extract values for the two-loop cusp anomalous dimension,
\begin{eqnarray}
A^{(2)}&=&\left(-{268\over 9}+{4\over 3}\pi^2\right)C_FC_A
+{80\over 9}C_F T_Fn_f\nn
&&+{32\over 9}C_FT_Fn_s\, .
\label{101}
\end{eqnarray}
The non-log part of the anomalous dimension is
\begin{eqnarray}
B^{(2)} &=& \left(4\pi^2-3-48\zeta(3)\right)C_F^2\nn
&&+\left(-\frac{961}{27}-\frac{11\pi^2}{3}+52 \zeta(3)\right) C_F C_A\nn
&&+\left(\frac{260}{27}+\frac{4\pi^2}{3}\right)C_FT_Fn_F\nn
&&+\left(\frac{167}{27}+\frac{\pi^2}{3}\right)C_F T_F n_s\, .
\label{102}
\end{eqnarray}
The log part of the matching at $M$ for equal Higgs and gauge boson masses is
\begin{eqnarray}
D_1^{(2)} &=&\left[{112\over 27}
+{4\over 9}\pi^2\right]C_FT_Fn_f + \left[-{782\over 27}
-\frac{20}{3}\zeta(3)\right.\nn
&&\left.+5\sqrt{3}\pi
+\frac{26}{\sqrt{3}}{\rm Cl}_2\left({\pi\over 3}\right)\right]C_F\,.
\label{103}
\end{eqnarray}
where the Clausen function is
\begin{eqnarray}
{\rm Cl}_2\left(x\right)&=&\sum_1^\infty \frac{\sin nx}{n^2}\, .
\end{eqnarray}

The anomalous dimension for $Q > \mu > M$ is independent of infrared physics, such as spontaneous symmetry breaking and the gauge boson mass, and so can be written in terms of group invariants such as $C_F$ and $C_A$. The expressions for $A^{(2)}$ and $B^{(2)}$ hold in a gauge theory with fermion and scalar fields in arbitrary representations. 

\begin{figure}
\begin{center}
\includegraphics[width=4cm]{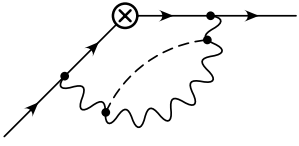}
\end{center}
\caption{\label{fig:ssb} A graph whose group-theoretic factor cannot be written
in terms of invariants such as $C_F$ and $C_A$.}
\end{figure}

The matching $D_1^{(2)}$ depends on the gauge boson masses, and is only valid in a $SU(2)$ gauge theory with scalars in the fundamental representation. The expression Eq.~(\ref{103}) has a $C_F T_F n_F$ term from fermion loop corrections to the gauge boson propagator, and a $C_F$ term. The $C_F$ term arises from scalar loop corrections to the gauge boson propagator, as well as graphs such as Fig.~\ref{fig:ssb} which arise due to spontaneous symmetry breaking.
The group theory invariant for Fig.~\ref{fig:ssb} depends on the pattern of symmetry breaking, and cannot be written in terms of $SU(2) \times U(1)$ invariants. Jantzen and Smirnov~\cite{js} have therefore explicitly used the group theory factors for a broken $SU
(2)$ theory in evaluating these contributions, and we follow their convention here. Furthermore, Ref.~\cite{js} computed the two-loop graphs only for $M_H=M_W$, and so Eq.~(\ref{103}) is only valid for equal Higgs and gauge boson masses.

The three-loop cusp anomalous dimension is known in a theory without scalar fields~\cite{MVV}
\begin{eqnarray}
\Gamma^{(3)}&=&C_F\Biggl[\left(-{245\over 3}+{268\over 27}\pi^2
-{44\over 3}\zeta(3)-{22\over 45}\pi^4\right)C_A^2\nn
&&+\left({836\over 27}-{80\over 27}\pi^2
+{112\over 3}\zeta(3)\right)C_AT_Fn_f
\nn
&&+\left({110\over 3}-{32}\zeta(3)\right)C_FT_Fn_f+
{32\over 27}(T_Fn_f)^2\Biggr]\,.\nn
\label{3loopgam}
\end{eqnarray}
so $A^{(3)}=2\Gamma^{(3)}$ is known, neglecting scalar contributions. These missing scalar contributions are expected to make small corrections to $\Gamma^{(3)}$. The scalar term contributes 7\% to the two-loop cusp anomalous dimension $A^{(2)}$.

Our one-loop computation combined with the known two-loop cusp anomalous dimension sums the LL and NLL series for the Sudakov form factor. The NNLL series requires $A^{(3)}$ which is known excluding Higgs contributions, $B^{(2)}$ which is known (Eq.~(\ref{102})), $D_1^{(2)}$ which is known for $M_H=M$, and $C^{(1)}, D_0^{(1)}$ which are known (Table~\ref{tab:results}). For electroweak applications, the LL and NLL are more than adequate for precision studies. 

\section{Proof that $D$ is linear in $\lQ$}
\label{sec:linear}

The general functional form of the $n$-collinear graphs is $\exp F(a(\mu),\lM,\lnii,\lnm)$, where $\lM=\log M^2/\mu^2$, $\lnii=\log \nu_2^2/\mu^2$ and $\lnm=\log \nu_1^-/p_1^-$. Using $\nu_1^-=\nu_1^2/p_1^+$ and $Q^2=p_1^+p_2^-$, this can be rewritten as $\exp F(a(\mu),\lM,\lnii,\lni-\lQ)$, where $\lQ=\log Q^2/\mu^2$. Similarly, the $\bar n$-collinear graphs have the functional form $\exp G(a(\mu),\lM,\lni,\lnii-\lQ)$. The sum of all the collinear graphs is the product $\exp(F+G)$, because the $n$ and $\bar n$-collinear graphs factor. These graphs give the matching coefficient $\exp D$, since the ultrasoft graphs vanish on-shell, so that $D$ has the additive form
\begin{eqnarray}
D(a(\mu),\lM,\lQ) &=& F(a(\mu),\lM,\lnii,\lni-\lQ) \nn
&&+ G(a(\mu),\lM,\lni,\lnii-\lQ)\, . \nn
\label{46a}
\end{eqnarray}
The $\lni,\lnii$ dependence cancels, since $D$ is independent of $\nu_{1,2}$.\footnote{There are also wavefunction contributions to $D$. These are independent of $Q$.}

Equation~(\ref{46a}) implies that $D(a(\mu),\lM,\lQ)$ is linear in $\lQ$. The proof is as follows:
Differentiating Eq.~(\ref{46a}) with respect to $\lni$ and $\lQ$, with respect to $\lnii$ and $\lQ$, and with respect to $\lni$ and $\lnii$ give
\begin{eqnarray}
0 &=& \partial_1 \partial_1 F + \partial_{2}\partial_{1} G\, , \nn
0 &=& \partial_{1}\partial_{2} F + \partial_{2}\partial_{2} G\, , \nn
0 &=& \partial_{2}\partial_{1}F + \partial_{2}\partial_{1} G\, ,
\label{47}
\end{eqnarray} 
where $\partial_{1,2}$ is the derivative with respect to $\mathsf{L}_{1,2}$.
The second derivative of $D$ with respect to $\lQ$ is
\begin{eqnarray}
\frac{\partial D}{\partial \lQ^2} &=& \partial_1 \partial_1 F + \partial_2\partial_2 G \nn 
&=& -\partial_2 \partial_1 G - \partial_1 \partial_2 F \nn 
&=& -\partial_1 \partial_2 \left(F+G\right)\nn
&=& -\partial_1 \partial_2 D =0\, ,
\label{48}
\end{eqnarray} 
using Eq.~(\ref{47}) and the commutation of partial derivatives, $\partial_3\partial_2=\partial_2\partial_3$. Thus $D$ can be at most linear in $\lQ$.  Equation~(\ref{46a}) is  $D$ before the addition of renormalization counterterms. The finite part shows that the matching correction is linear in $\lQ$, justifying the form Eq.~(\ref{34}) used earlier. The infinite part shows that the SCET anomalous dimension is linear in $\lQ$, and so gives another proof of this known result~\cite{dis,Bauer:2003pi}.

\section{Consistency Conditions}
\label{sec:cc}

There are consistency conditions on matching coefficients and anomalous dimensions which follow from the structure of the effective theory. Consider the matching of an operator between a high energy theory and a low energy theory at some scale $\mu$. The operator coefficients are $c_{h,l}(\mu)$, with anomalous dimensions $\gamma_{h,l}(\mu)$ in the two theories, $\mu {\rm d}c_{h,l}/{\rm d}\mu=\gamma_{h,l} c_{h,l}$. Assume that there is a multiplicative matching coefficient $X(\mu)$ between the two theories, so that $c_l(\mu)=X(\mu) c_h(\mu)$. The matching scale $\mu$ is arbitrary, so one gets the constraint
\begin{eqnarray}
\mu \frac{ {\rm d}}{ {\rm d} \mu} \log c_l &=& \mu \frac{ {\rm d}}{ {\rm d} \mu} \log X + \mu \frac{ {\rm d}}{ {\rm d} \mu} \log c_h\, ,
\label{39}
\end{eqnarray}
which gives the relation
\begin{eqnarray}
\gamma_l-\gamma_h &=& \mu \frac{ {\rm d}}{ {\rm d} \mu} \log X\, ,
\label{36}
\end{eqnarray}
between the matching coefficient and the anomalous dimensions in the two theories. In the Sudakov problem, applying Eq.~(\ref{36}) to the matching between the full theory and SCET gives
\begin{eqnarray}
\gamma-\gamma_F &=& \mu \frac{ {\rm d}}{ {\rm d} \mu} C\, ,
\label{37}
\end{eqnarray}
and applying it to the matching when the gauge boson is integrated out gives
\begin{eqnarray}
0-\gamma &=& \mu \frac{ {\rm d}}{ {\rm d} \mu} D\, ,
\label{38}
\end{eqnarray}
since the theory below $M$ is a free theory, and so has zero anomalous dimension.

The anomalous dimension in the full theory and SCET have the form $\gamma_F(a(\mu))$ and $\gamma = - A(a(\mu)) \lQ + B(a(\mu))$, respectively, where we have used the result that $\gamma$ is linear in $\lQ$ to all orders~\cite{dis}. The matching coefficient $C$ at the high scale $Q$ is independent of the low-energy scales such as $M$, and has the form $C(a(\mu),\lQ)$. The matching coefficient $D$ at $\mu \sim M$ can depend on both $Q$ and $M$, since the SCET field labels depend on $Q$. As a result, $D$ has the form $D(a(\mu),\lM,\lQ)$. Any dependence on $Q^2/M^2$ can be converted into dependence on $\lM$ and $\lQ$ using $Q^2/M^2=\exp(\lQ-\lM)$.

Equation~(\ref{37}) gives
\begin{eqnarray}
A(a(\mu)) \lQ - B(a(\mu))+\gamma_F(a(\mu)) &=&  2  \frac{\partial C}{\partial \lQ} -
\frac{\partial C}{\partial a} \beta_a(a)\, . \nn
\label{40}
\end{eqnarray}
Writing $C$ as an expansion in $\lQ$
\begin{eqnarray}
C(a(\mu),\lQ) &=& \sum_{n=0}^\infty C_n(a(\mu)) \lQ^n\, ,
\label{41}
\end{eqnarray}
gives the consistency conditions
\begin{eqnarray}
\gamma_F(a)- B(a) &=& 2 C_1(a) - \frac{\partial C_0}{\partial a} \beta_a(a)\, ,\nn
A(a) &=& 4 C_2(a) - \frac{\partial C_1}{\partial a} \beta_a(a)\, ,\nn
2 n  C_n(a) &=&  \frac{\partial C_{n-1}}{\partial a} \beta_a(a)\, ,\qquad n \ge 3\, ,
\label{42}
\end{eqnarray}
which determine $C_n$, $n>0$ in terms of $C_0$, $A$, $B$ and $\gamma_F$, and
are satisfied by the one-loop values in Table~\ref{tab:results}. The matching coefficient at $\mu=Q$ is $C_0(a(Q))$.

Equation~(\ref{38}) applied to the matching at $M$, gives
\begin{eqnarray}
A(a(\mu)) \lQ - B(a(\mu)) &=&  \frac{\partial D}{\partial a} \beta_a(a)- 2  \frac{\partial D}{\partial \lQ} -  2  \frac{\partial D}{\partial \lM}\, .\nn
\label{43}
\end{eqnarray}
Using Eq.~(\ref{34}), and equating powers of $\lQ$ gives
\begin{eqnarray}
 - B(a(\mu)) &=&  \frac{\partial D_0}{\partial a} \beta_a(a)- 2  D_1  -  2  \frac{\partial D_0}{\partial \lM}\, ,\nn
A(a(\mu))  &=&  \frac{\partial D_1}{\partial a} \beta_a(a) -  2  \frac{\partial D_1}{\partial \lM}\, .
\label{50}
\end{eqnarray}
Writing $D_{i}$, $i=0,1$ as an expansion in $\lM$
\begin{eqnarray}
D_{i}(a(\mu),\lM) &=& \sum_{n=0}^\infty D_{i,n}(a(\mu)) \lM^n\, ,
\label{51}
\end{eqnarray}
gives the consistency conditions
\begin{eqnarray}
- B &=&  \frac{\partial D_{0,0}}{\partial a} \beta_a-2 D_{1,0}-2 D_{0,1}\, , \nn
2n D_{0,n} &=&  \frac{\partial D_{0,n-1}}{\partial a} \beta_a-2 D_{1,n-1}\, ,\qquad
 n\ge 2\, ,\nn
A &=&  \frac{\partial D_{1,0}}{\partial a} \beta_a-2 D_{1,1}\, ,\nn
2 n  D_{1,n}(a) &=&  \frac{\partial D_{1,n-1}}{\partial a} \beta_a\, ,\qquad n \ge 2\, ,
\label{52}
\end{eqnarray}
which determine $D_{i,n}$, $n>0$ in terms of $D_{i,0}$, $A$ and $B$, and
are satisfied by the one-loop values in Table~\ref{tab:results}. The matching coefficient at $\mu=M$ is $D_{0,0}(a(M))+D_{1,0}(a(M)) \log Q^2/M^2$.

\section{Massive Particles}
\label{sec:mass}

The calculations so far have been performed for external particles with masses $m_{1,2}$ much smaller than the gauge boson mass. In this section, we extend the results to massive external particles.  For fermions, we will assume that the theory is vectorlike, so that the fermion mass arises from a gauge invariant mass term $- m \bar \psi \psi$. The standard model, a chiral gauge theory, where masses arise from Higgs couplings due to spontaneous symmetry breaking, is discussed in Sec.~\ref{sec:stdmodel}.

\subsection{$Q \gg m_2 \gg M \gg m_1$}
\label{sec:massA}

Consider first the case where one particle has mass $m_2$, with $Q \gg m_2 \gg M$, and the other particle has mass $m_1$ much smaller than $M$. For definiteness, the outgoing particle is taken to be heavier than $M$, and the incoming one lighter than $M$, but the results are symmetric under $1\leftrightarrow 2$. The Sudakov form-factor can be computed using a sequence of effective field theories. 

One first matches from the full theory to SCET with a massive particle at $\mu \sim Q$ and uses the same set of operators listed in Eq.~(\ref{19}) except $\xi_{n,p_2}$ is now a $n$-collinear SCET field with mass $m_2$ as in Ref.~\cite{llw}.  This matching is independent of scales much smaller than $Q$, such as $m_1$, $m_2$ and $M$, and thus remains the same as in Table~\ref{tab:results}. The second step is to run the operator in the effective theory from $Q$ to $m_2$. The anomalous dimension $\gamma$ is also independent of low mass scales and again gives the same result as the massless case.  At the scale $m_2$, one switches from SCET to a new effective theory in which the massive particle is described by a heavy quark effective theory (HQET) field $h_{v_2}$, with a velocity $v_2$, with $v_2^2=1$~\cite{book}.  The other (massless) particle is still described by a SCET field. The fermion vector current, for example, is now given by $\bar h_{v_2} \gamma^\mu W^\dagger_{\bar n} \xi_{\bar n,p_1}$, instead of Eq.~(\ref{19}), and similarly for the other operators. The HQET field, $h_{v_2}$, does not transform under collinear gauge transformations; therefore, there is no factor analogous to the $W_{\bar n}^\dagger$ Wilson line that goes along with the $\xi_{\bar n}$ field. $h_v$ still couples to ultrasoft gluons. One can make an additional field redefinition which eliminates the ultrasoft gluon coupling to $h_v$ and introduces a Wilson line in the $v_2$ direction~\cite{fhms}. Both methods give the same on-shell matrix elements.

\begin{figure}
\begin{center}
\includegraphics[width=4cm]{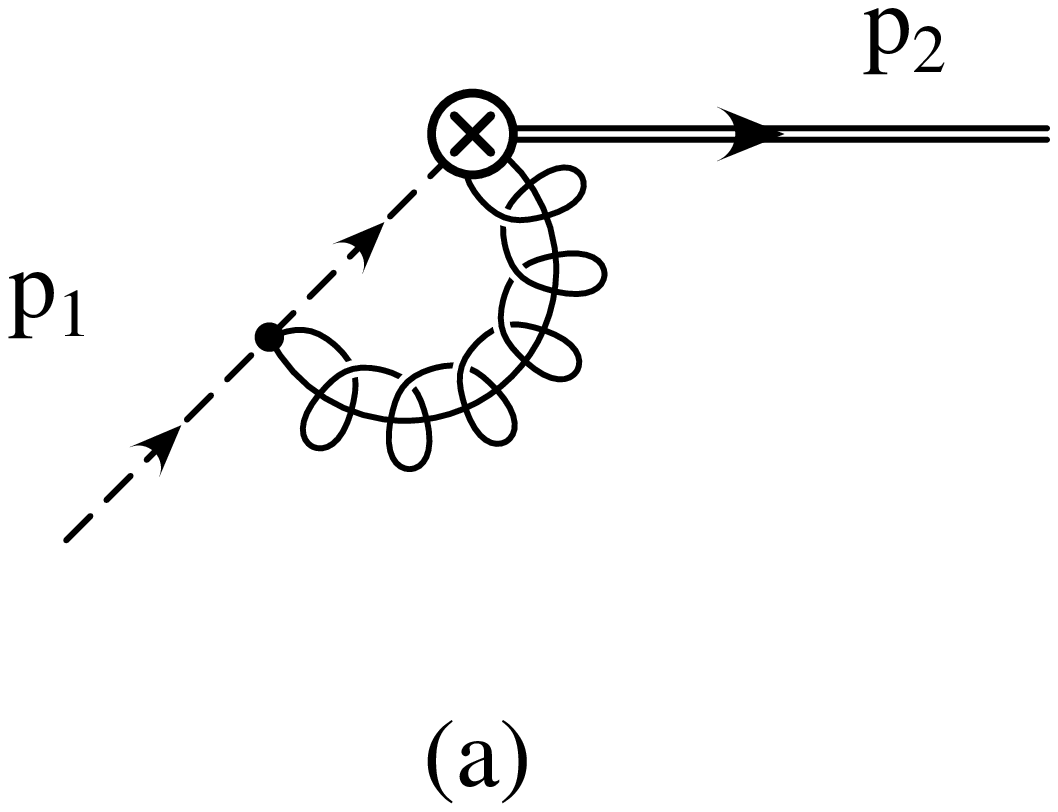}
\includegraphics[width=4cm]{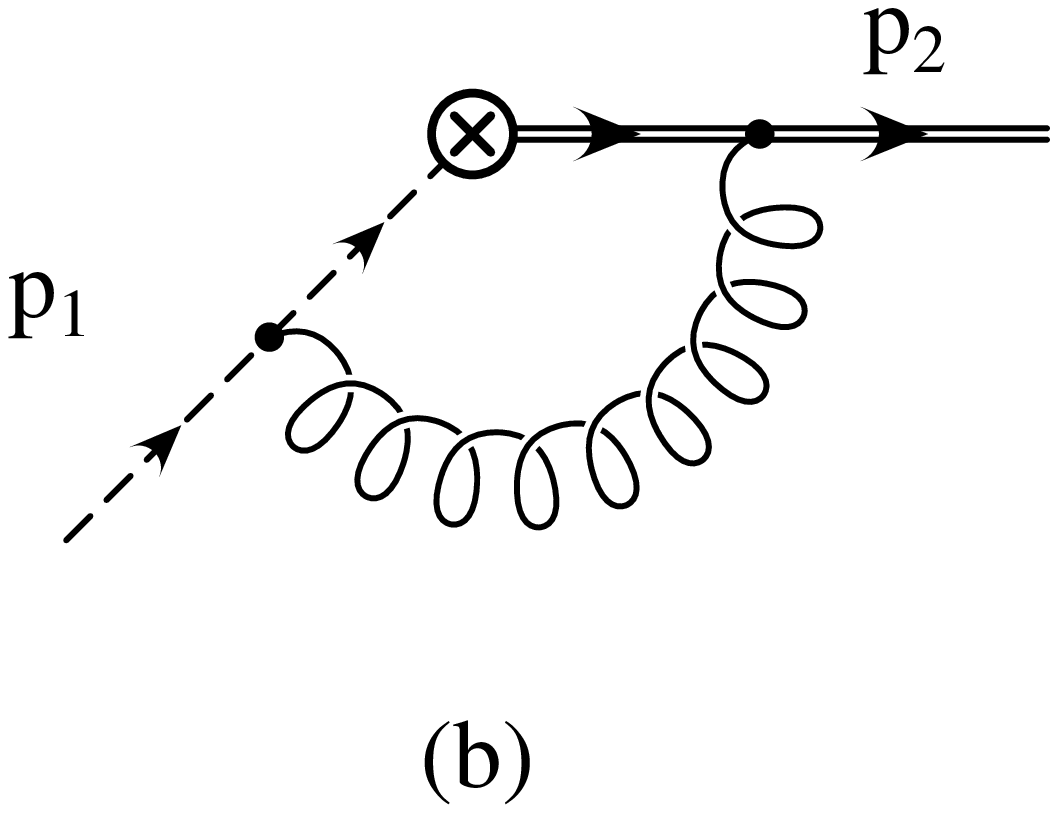}
\end{center}
\caption{\label{fig:hqet} SCET graphs for the matrix element of $\widetilde\mathcal{O}$. The dotted lines are SCET propagators, and represent either fermions or scalars. The double lines are HQET propagators.}
\end{figure}

The matching condition at $m_2$ is given by the difference between the vertex graphs in Fig.~\ref{fig:scet} and Fig.~\ref{fig:hqet},
and the wavefunction graphs, evaluated at $\mu=m_2$,  i.e.\ between graphs where $\xi_{n,p_2}$ and $h_{v_2}$ are used for particle~2. Note that there are three vertex graphs in the theory above $m_2$, and only two graphs in the theory below $m_2$, because there is no collinear Wilson line associated with $h_{v_2}$. The graphs in the theory above $m_2$ are evaluated with the gauge boson mass set to zero (since $m_2 \gg M$) at the on-shell point $p_2^2=m_2^2$. The graphs in the effective theory are evaluated with $M \to 0$ at the onshell point $k_2 \cdot v_2=0$, where $k_2$ is the residual momentum of particle $2$. Graphs Fig.~\ref{fig:scet}b and Fig.~\ref{fig:hqet}a are equal, since the $\bar n$ collinear interactions do not depend on whether the other field at the vertex is $\bar h_v$ or $\bar \xi_{n,p_2}$. They cancel in the matching computation. The ultrasoft graphs Fig.~\ref{fig:scet}c and Fig.~\ref{fig:hqet}b, the $\xi_{\bar n,p_1}$ wavefunction graph, and the HQET wavefunction graph all vanish on-shell, so the matching computation is given by  Fig.~\ref{fig:scet}a and the on-shell wavefunction graph for $\xi_{n,p_2}$. The vertex graph Fig.~\ref{fig:scet}a does not need an analytic regulator, and gives (for $\mathcal{O}=\bar \psi \gamma^\mu \psi$)
\begin{eqnarray}
I_n &=&a C_F \gamma^\mu \Biggl[ \frac{1}{\epsilon^2} +\frac{1}{\epsilon}\biggl(2 -  \log \frac{m_2^2}{\mu^2}
 \biggr) \nn
 &&+\frac12\log^2 \frac{m_2^2}{\mu^2}-2 \log \frac{m_2^2}{\mu^2}
 +\frac{\pi^2}{12}+4
\Biggr]\, .
\label{58}
\end{eqnarray}
The wavefunction correction for a massive fermion is
\begin{eqnarray}
\delta z&=& a C_F \left[ \frac{1}{\eUV}+\frac{2}{\eIR}+4 - 3 \log \frac{m_2^2}{\mu^2}\right]\, .
\label{56}
\end{eqnarray}
Combining Eq.~(\ref{58}) and Eq.~(\ref{56}) gives the multiplicative matching condition $\exp R$, $R=R^{(1)} a+ \ldots$,  at $\mu$ of order $m_2$,
\begin{eqnarray}
R &=& a C_F \Biggl[ \frac12\lmii^2-\frac12\lmii
 +\frac{\pi^2}{12}+2\Biggr]\, ,
\label{60}
\end{eqnarray}
where $\lmii \equiv \log m_2^2/\mu^2$. The other cases are evaluated in a similar manner,
and the results are summarized in Table~\ref{tab:mresults}. The wavefunction correction for a massive scalar, which is needed for the last three rows, is
\begin{eqnarray}
\delta z&=& a C_F \left[ -\frac{2}{\eUV}+\frac{2}{\eIR}\right]\, .
\label{57}
\end{eqnarray}
\begin{table*}
\begin{eqnarray*}
\begin{array}{|c|c|c|c|c|}
\hline
\mathcal{O} & R^{(1)}/C_F & \gamma_2^{(1)}/C_F &S^{(1)}/C_F \\
\hline
\bar \psi_2 \Gamma \psi_1 & \frac12\lmii^2-\frac12\lmii  +\frac{\pi^2}{12}+2 
& -5 - 2 \lmii + 4 \lQ & -\frac52 \lM +\frac94-\frac{5 \pi^2}{12}-\frac12  \lM^2 -  \lM \left( \lmii-2\lQ \right)\\[5pt]
\phi^\dagger_2 \phi_1, i(\phi^\dagger_2 D^\mu \phi_1-D^\mu\phi^\dagger_2  \phi_1) & \frac12\lmii^2-\lmii  +\frac{\pi^2}{12}+2 & -6 - 2 \lmii + 4 \lQ & -3 \lM +\frac74-\frac{5 \pi^2}{12}-\frac12  \lM^2 -  \lM \left( \lmii-2\lQ \right) \\[5pt]
\bar \psi_2 \phi_1 & \frac12\lmii^2-\frac12\lmii  +\frac{\pi^2}{12}+2 & -6 - 2 \lmii + 4 \lQ & -3 \lM +\frac74-\frac{5 \pi^2}{12}-\frac12  \lM^2 -  \lM \left( \lmii-2\lQ \right) \\[5pt]
\phi^\dagger_2 \psi_1 & \frac12\lmii^2-\lmii  +\frac{\pi^2}{12}+2 & -5 - 2 \lmii + 4 \lQ & -\frac52 \lM +\frac94-\frac{5 \pi^2}{12}-\frac12  \lM^2 -  \lM \left( \lmii-2\lQ \right) \\[5pt]
\hline
\end{array}
\end{eqnarray*}
\caption{\label{tab:mresults} One-loop results for $Q > m_2 > M >m_1$. $R$ is the matching coefficient at $\mu \sim m_2$, $\gamma_2$ is the anomalous dimension between $m_2$ and $M$, and $S$ is the matching coefficient at $\mu \sim M$. The results only depend on whether the light particle is a fermion or scalar.}
\end{table*}

\begin{figure}
\begin{center}
\includegraphics[width=4cm]{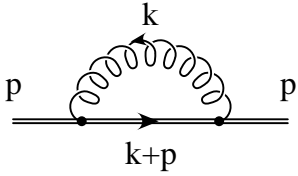}\
\end{center}
\caption{\label{hqet:prop} HQET propagator Correction  }
\end{figure}

The remaining steps are the evaluation of the anomalous dimension in the region $M < \mu < m_2$, and the matching condition $\exp S$ at the scale $M$. These can be computed by evaluating the graphs in Fig.~\ref{fig:hqet} and the wavefunction corrections, at the on-shell point $k_2=0$, $p_1^2=0$, and keeping the gauge boson mass non-zero.  The graphs need to be regulated using an analytic regulator. The $\xi_{\bar n,p_1}$ is regulated as in Eq.~(\ref{22}) for the collinear propagator. The Wilson line in Fig.~\ref{fig:hqet}a is regulated using Eq.~(\ref{23}), as is the massless particle propagator in Fig.~\ref{fig:hqet}b. The new feature is the HQET propagator for particle 2, which is regulated using
\begin{eqnarray}
\frac{1}{k_2 \cdot v_2} &\to&
\frac{ (-\nu_{2H})^{\delta_2}}
{\left[k_2 \cdot v_2 \right]^{1+\delta_2}}\: .
\label{61}
\end{eqnarray}
Taking the HQET limit of the particle 2 propagator,
\begin{eqnarray}
\frac{ (-\nu_2^2)^{\delta_2}}{\left[(m_2 v_2+k_2)^2-m_2^2\right]^{1+\delta_2}} &\to& \frac{ (-\nu_2^2)^{\delta_2}}
{\left[2 m_2 k_2 \cdot v_2 \right]^{1+\delta_2}}\: .
\label{59}
\end{eqnarray}
 and comparing Eq.~(\ref{61}) with Eq.~(\ref{22}), we see that $\nu_{2H}=\nu_2^2/(2m_2)$. Figure~\ref{fig:hqet}a is given by the $\bar n$-collinear graph evaluated earlier, Eq.~(\ref{25b}). Figure~\ref{fig:hqet}b is
\begin{eqnarray}
&& -a C_F \gamma^\mu\left(\frac{2 \nu_H}{M}\right)^{\delta_2}\left(\frac{\nu_1^-}{M}\right)^{\delta_1}
\left( \frac{\mu^2}{M^2} \right)^\epsilon \nn
&&\times  \frac{\Gamma(\delta_2/2-\delta_1/2)
\Gamma(\epsilon+\delta_1/2+\delta_2/2)}{\Gamma(1+\delta_2)}\, .
\label{62}
\end{eqnarray}
The $\bar n$-collinear wavefunction graph is Eq.~(\ref{31}), and the HQET propagator correction, Fig.~\ref{hqet:prop}
is
\begin{eqnarray}
a C_F \left[ 2M\pi + (k_2 \cdot v_2)\left(-\frac{2}{\epsilon}
+2 \log\frac{M^2}{\mu^2}\right)\right]\, .
\label{63}
\end{eqnarray}
Eq.~(\ref{63}) gives a contribution to the heavy quark residual mass term, $\delta m = - 2 a C_F M \pi$, and a wavefunction contribution listed in Table~\ref{tab:wave}. The shift in the heavy quark mass is non-analytic in the gauge boson mass-squared. Such non-analytic contributions occur in mass corrections to particles with $k \cdot v$ propagators~\cite{chpt,hoang}. They arise from loop integrals which diverge as an odd power of $k$. Such integrals are finite, but non-analytic, in dimensional regularization.

Adding Eqs.~(\ref{25b}) and Eq.~(\ref{62}), expanding in $\delta_i$, and subtracting the wavefunction corrections due to the heavy fermion, Eq.~(\ref{63}), and the collinear fermion, Eq.~(\ref{31}), gives
\begin{eqnarray}
&=& a C_F 
\Biggl[\frac1{\epsilon^2}+\frac1{\epsilon}\left(\frac52 + 2 \log \frac{\mu m_2}{Q^2}\right)+\frac52 \log \frac{\mu^2}{M^2}\nn
&&+\frac94-\frac{5 \pi^2}{12}-\frac12  \log^2 \frac{\mu^2}{M^2} +2 \log \frac{\mu^2}{M^2}\log \frac{\mu m_2}{Q^2}
 \Biggr].\nn
 \label{64}
\end{eqnarray} 
To obtain Eq.~(\ref{64}) we have used $\nu_{2H}=\nu_2^2/(2m_2)$, $\nu_1^-=\nu_1^2/p_1^+$, $\nu_2^+=\nu_2^2/p_2^-$, and $Q^2=m_2 p_1^+$. The $1/\epsilon$ coefficient multiplied by $-2$ gives the anomalous dimension
\begin{eqnarray}
\gamma_2 &=& a C_F \left(-5 - 2 \lmii + 4 \lQ \right)\, ,
\label{65}
\end{eqnarray} 
and the finite part gives the matching correction
\begin{eqnarray}
S  &=& a C_F 
\Biggl[-\frac52 \lM +\frac94-\frac{5 \pi^2}{12}\nn
&&-\frac12  \lM^2 -  \lM \left( \lmii-2\lQ \right)
 \Biggr]\, .
 \label{66}
\end{eqnarray} 
The other cases are computed similarly, and are given in Table~\ref{tab:mresults}.

\begin{table*}
\begin{eqnarray*}
\begin{array}{|c|c|c|c|c|}
\hline
\mathcal{O} & T^{(1)}/C_F & \gamma_3^{(1)}/C_F &U^{(1)}/C_F \\
\hline
\bar \psi_2 \Gamma \psi_1 & \frac12\lmi^2-\frac12\lmi  +\frac{\pi^2}{12}+2 
& 4\left[ w r(w)-1\right] & 2 \left[ w r(w)-1\right]\lM \\[5pt]
\phi^\dagger_2 \phi_1, i(\phi^\dagger_2 D^\mu \phi_1-D^\mu\phi^\dagger_2  \phi_1) & \frac12\lmi^2-\lmi  +\frac{\pi^2}{12}+2 & 4\left[ w r(w)-1\right] &  2 \left[ w r(w)-1\right]\lM \\[5pt]
\bar \psi_2 \phi_1 & \frac12\lmi^2-\lmi  +\frac{\pi^2}{12}+2 & 4\left[ w r(w)-1\right] &  2 \left[ w r(w)-1\right]\lM \\[5pt]
\phi^\dagger_2 \psi_1 & \frac12\lmi^2-\frac12 \lmi  +\frac{\pi^2}{12}+2 & 4\left[ w r(w)-1\right]&  2 \left[ w r(w)-1\right]\lM \\[5pt]
\hline
\end{array}
\end{eqnarray*}
\caption{\label{tab:mmresults} One-loop results for $Q > m_2 > m_1 >M$. $T$ is the matching at $m_1$, $\gamma_3$ is the anomalous dimension between $m_1$ and $M$, and $U$ is the matching at $M$. $T$ only depends on whether the light particle is a scalar or a fermion}
\end{table*}

The terms which contribute to the final result are summarized schematically as:
\begin{eqnarray}
\begin{array}{ccccccc}
C & \gamma_1 & R & \gamma_2 & S \\
Q & \longrightarrow &  m_2 &  \longrightarrow & M 
\end{array}
\end{eqnarray}

\subsection{$Q \gg m_2 \gg m_1 \gg M$}
\label{sec:massB}

The second case we consider is where both particles have mass between $Q$ and $M$. For definiteness, we choose $m_2>m_1$. The Sudakov form-factor can be computed using a sequence of matching and running steps. The matching $\exp C$ at $Q$, the anomalous dimension $\gamma_1$ between $Q$ and $m_2$, the matching $\exp R$ at $m_2$, and the anomalous dimension $\gamma_2$ between $m_2$ and $m_1$ are all independent of the lower scales $m_1$ and $M$, and have the same values as in Tables~\ref{tab:results}, \ref{tab:mresults}. 

\begin{figure}
\begin{center}
\includegraphics[width=4cm]{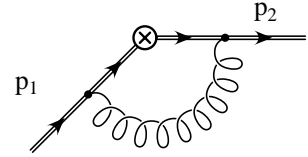}\
\end{center}
\caption{\label{fig:dhqet} Vertex Correction in the theory below $m_{1,2}$. }
\end{figure}

The new feature is the matching condition $\exp T$ at the lower particle mass $m_1$. The graphs in the theory above $m_1$ are shown in Fig.~\ref{fig:hqet}. In the theory below $m_1$, the SCET field $\xi_{\bar  n,p_1}$ for particle 1 is replaced by the HQET field $h_{v_1}$. The fermion vector current, for example, is now given by $\bar h_{v_2} \gamma^\mu h_{v_1}$ instead of $\bar h_{v_2} \gamma^\mu W^\dagger_{\bar n} \xi_{\bar  n,p_1}$. The vertex correction in the theory below $m_1$ is shown in Fig.~\ref{fig:dhqet}. 
There is only one vertex graph instead of two, because there is no collinear Wilson line $W$ associated with the HQET field $h_{v_1}$. The matching condition is given by computing the difference of graphs Fig.~\ref{fig:hqet} and Fig.~\ref{fig:dhqet} on-shell, and setting all scales less than $m_1$ to zero. The only non-zero contribution is from Fig.~\ref{fig:hqet}a and the $\bar n$-collinear wavefunction renormalization graph. These are the same graphs that contribute to the matching condition at $m_2$, so $T$ is given by $R$ with $m_2 \to m_1$,\footnote{$T$ depends on whether the particle being integrated out is a fermion or a scalar.}\ and is tabulated in Table~\ref{tab:mmresults}.

The remaining quantities needed are the anomalous dimension $\gamma_3$ between $m_1$ and $M$, and the matching condition $\exp U$ at $M$. These can be computed from the graph in Fig.~\ref{fig:dhqet} evaluated on-shell, but now with the gauge boson mass $M$ included. The graph gives
\begin{eqnarray}
a\, w r(w) \left[-\frac{2}{\epsilon}+2 \log \frac{M^2}{\mu^2} \right]\, ,
\label{66a}
\end{eqnarray}
where $w=v_2\cdot v_1$ and
\begin{eqnarray}
r(w) &=&\frac{ \log \left(w + \sqrt{w^2-1}\right)}{\sqrt{w^2-1}}\, ,
\label{67}
\end{eqnarray}
is the factor which occurs in the velocity-dependent anomalous dimension in HQET~\cite{book}. Including the heavy  quark wavefunction correction, Eq.~(\ref{63}), gives the anomalous dimensions and matching coefficient listed in Table~\ref{tab:mmresults}. In the high energy  limit $Q^2 \sim 2 m_1 m_2 w$,
\begin{eqnarray}
w r(w) \sim \log(2 w)\, .
\label{68}
\end{eqnarray}

The terms which contribute to the final result are:
\begin{eqnarray}
\begin{array}{ccccccc}
C & \gamma_1 & R & \gamma_2 & T & \gamma_3 & U \\
Q & \longrightarrow &  m_2 &  \longrightarrow & m_1 & \longrightarrow & M 
\end{array}
\end{eqnarray}

\subsection{$ Q \gg m_2=m_1 \gg M$}
\label{sec:massC}

The third case we consider is where the two particles are degenerate, with $m_2=m_1=m$, and $Q \gg m \gg M$. This can be computed using the results already derived. The matching $\exp C$ at $Q$ and the running $\gamma_1$ between $Q$ and $m$ is the same as in Table~\ref{tab:results}. The matching at $m$ is given by switching both particles from SCET to HQET simultaneously. This is just the sum of the matching coefficients at $m_2$ and $m_1$ computed previously, so the matching condition is
$\exp\left[R+T\right]$, with $m_1=m_2=m$, where $R$ and $T$ are given in Tables~\ref{tab:mresults} and \ref{tab:mmresults}, respectively. The anomalous dimension between $m$ and $M$, and the matching at $M$ are given by $\gamma_3$ and $U$ in Table~\ref{tab:mmresults}.

The terms which contribute to the final result are:
\begin{eqnarray}
\begin{array}{ccccccc}
C & \gamma_1 & R + T & \gamma_3 & U \\
Q & \longrightarrow &  m_2=m_1  & \longrightarrow & M 
\end{array}
\end{eqnarray}

\subsection{$ Q \gg m_2  \sim M \gg m_1 $}
\label{sec:massD}

It is also useful to derive results where the particle masses and gauge boson masses are not widely separated from each other. In this case, it is more important to include the full $m/M$ dependence, rather than sum high order $\alpha \log m/M$ terms, which are no longer very large. In the standard model, this situation arises for the $W$, $Z$ and $t$, which have masses which are not sufficiently widely separated that electroweak logarithms need to be summed. If two (or more) particle masses are not widely separated, one can make a transition to a new EFT by integrating out both particles at a common scale $\mu$, rather than integrating them out sequentially. The results for the various cases are summarized in this and the following subsections. The difference from previous results is that one has to include all the relevant masses in the particle propagators.  For example, for the case studied in this subsection, $m_2 \sim M$,  one also includes $m_2$ in the denominator of Eq.~(\ref{21}). The integrals now depend on a dimensionless parameter, the ratio of particle masses $m_2^2/M^2$. 

If $Q \gg m_2  \sim M \gg m_1 $, then the matching at $Q$ and the running between $Q$ and $m_2 \sim M$, remain unchanged, and are given by $C$ and $\gamma_1$ in Table~\ref{tab:results}. At the scale $\mu$ of order $m_2\sim M$, one integrates out the gauge boson, and switches to a theory in which particle~2 is described by a HQET field.
In this matching, the $n$-collinear graph Eq.~(\ref{21}) with the analytic regulator is now
\begin{eqnarray}
I_n &=&-i g^2 \mu^{2\epsilon} C_F c(\mu) \int { {\rm d}^d k \over (2 \pi )^d} 
\frac{1}{k^2-M^2}{\slashed{\bar n}\  \over 2}  n^\alpha\nn
&\times&  \frac{\slashed{n} }{2} \frac{ (-\nu_2^2)^{\delta_2} \bar n \cdot (p_2-k) }{
\left[(p_2 - k )^2-m_2^2\right]^{1+\delta_2} } \gamma^\mu \frac{(-\nu_1^-)^{\delta_1}}{\left[- \bar n \cdot k\right]^{1+\delta_1}}\bar n_\alpha\ ,  \nn
\label{21m2}
\end{eqnarray} 
and is evaluated at the on-shell point $p_2^2=m_2^2$. The $\bar n$-collinear and ultrasoft integrals remain unchanged. We saw in Sec.~\ref{sec:massA} that the $\delta$ and $\nu$ dependence canceled between all the diagrams for the massless case. The cancellation must still hold when $I_n$ is evaluated with $m_2\not=0$, so that $aC_F f_F(z)=I_n(m_2)-I_n(m_2=0)$, $z=m_2^2/M^2$ has a finite limit independent of the analytic regulator, as can be verified by explicit computation. The result for $f_F(z)$ is given in Eq.~(\ref{fF}). The wavefunction renormalization is also modified, and the shift $h_F(z)$ is given in Eq.~(\ref{hF}). The matching condition $D$ is given by $D(m_2)=D(m_2=0) + a C_F \left( f_F(z)-h_F(z)/2\right)$ if the particle integrated out is a heavy fermion, where the massless value $D(m_2=0)$ is given in Table~\ref{tab:results}. Similarly, if the particle integrated out is a heavy scalar, the matching $D$ is  $D(m_2)=D(m_2=0) + a C_F \left( f_S(z)-h_S(z)/2\right)$ where the scalar functions are given in Eqs.~(\ref{fS},\ref{hS}).

The theory below $m_2\sim M$ is a theory in which particle~2 is described using a HQET field, and particle~1 by an SCET, with the massive gauge boson integrated out. In our toy example, this is a free theory.

Schematically, the terms are ($z=m_2^2/M^2$):
\begin{eqnarray}
\begin{array}{ccccccc}
C & \gamma_1 & D + a C_F(f(z)-h(z)/2) \\
Q & \longrightarrow &  m_2\sim M   
\end{array}
\end{eqnarray}

\subsection{$ Q \gg m_2  \gg M \sim m_1 $}
\label{sec:massE}

If $ Q \gg m_2  \gg M \sim m_1 $, the matching and running down to $M\sim m_1$ remains the same as in Sec.~\ref{sec:massA}. In the theory at $M \sim m_1$, particle~2 is described by an HQET field, and particle~1 by a $\bar n$ collinear field. The matching condition at $M \sim m_1$ would be given by $S$ if $m_1\to 0$. By the same arguments as above, the effect of $m_1$ is to modify the $\bar n$-collinear integral by a finite amount, so the matching is now $S+aC_F(f(z)-h(z)/2)$, with $z=m_1^2/M^2$, where $f,h$ are the fermion or scalar values Eqs.~(\ref{fF},\ref{hF}) or Eqs.~(\ref{fS},\ref{hS}), depending on the type of particle~1.

Schematically, the terms are ($z=m_1^2/M^2$):
\begin{eqnarray}
\begin{array}{ccccccc}
C & \gamma_1 & R & \gamma_2 &  S + a C_F(f(z)-h(z)/2) \\
Q & \longrightarrow &  m_2 & \longrightarrow & m_1 \sim M   
\end{array}
\end{eqnarray}

\subsection{$ Q \gg m_1  \sim m_2 \gg M $}
\label{sec:massF}

The situation is similar to $Q \gg m_1 =m_2 \gg M$ considered in Sec.~\ref{sec:massC}. 
The evolution down to $m_1\sim m_2$ is the same as for $m_i=0$. The $n$ and $\bar n$ collinear graphs at the scale $m_1 \sim m_2$ are independent of each other, so the matching is given by $R(m_2)+T(m_1)$ given in Tables~\ref{tab:mresults}, \ref{tab:mmresults}. Below $m_1 \sim m_2$, the computation reduces to that in Sec.~\ref{sec:massC}.

Schematically, the terms are:
\begin{eqnarray}
\begin{array}{ccccccc}
C & \gamma_1 & R +T & \gamma_3 &  U \\
Q & \longrightarrow &  m_1 \sim m_2 & \longrightarrow &  M   
\end{array}
\end{eqnarray}

\subsection{$ Q \gg m_1  \sim m_2 \sim M $}
\label{sec:massG}

The evolution to $\mu \sim m_1\sim m_2\sim M$ is the same as for the massless case. The matching at $\mu$ involves massive collinear propagators, each of which modifies the massless matching condition, so the matching is given by
\begin{eqnarray}
D(m_1,m_2) &=& D(m_1=m_2=0) +a C_F\left( f_2(z_2)-h_2(z_2)/2 \right)\nn
&&+aC_F \left(f_1(z_1)-h_1(z_1)/2\right)
\label{77}
\end{eqnarray}
where $z_i=m_i^2/M^2$, and $f_{1,2}$, $h_{1,2}$ are chosen to be $f_{F,S}$ and $h_{F,S}$ depending on whether the corresponding particle is a fermion or scalar. The massless value $D(m_1=m_2=0)$ is given in Table~\ref{tab:results}.

Schematically, the terms are ($z_i=m_i^2/M^2$):
\begin{eqnarray}
\begin{array}{ccccccc}
C & \gamma_1 & D+ aC_F(f(z_1)-h(z_1)/2)+aC_F(f(z_2)-h(z_2)/2) \\
Q & \longrightarrow &  m_1 \sim m_2 \sim  M   
\end{array}\nn
\end{eqnarray}

\section{Scalar corrections}
\label{sec:scalar}

In this section, we compute the scalar exchange corrections to the Sudakov form-factor. The graphs are the same as those for gauge exchange, with the gauge boson replaced by the scalar $\chi$, with mass $M_\chi$.  As for gauge bosons, one needs to include both collinear and ultrasoft fields for $\chi$ to represent collinear and ultrasoft $\chi$ particles. In the gauge boson results we removed an overall factor of $a=g^2/(16\pi^2)$. In the $\chi$-exchange graphs, we remove a factor of $h_1 h_2/(16\pi^2)$, where $h_{1,2}$ are the Yukawa couplings at the two vertices. Note that the coupling of $\chi$ to scalars $h_{\phi,i}$ has dimensions of mass, so the factor removed for operators such as $\phi^\dagger \phi$ is dimensionful. Unlike for gauge interactions, the wavefunction and vertex corrections can have different coupling constants.

\begin{figure}
\begin{center}
\includegraphics[width=4cm]{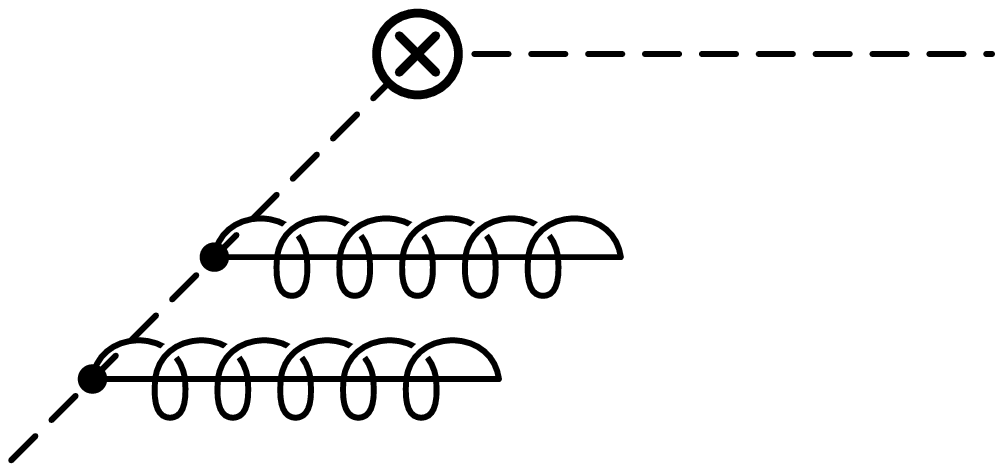}
\raise1cm\hbox{$\longrightarrow$}\includegraphics[width=4cm]{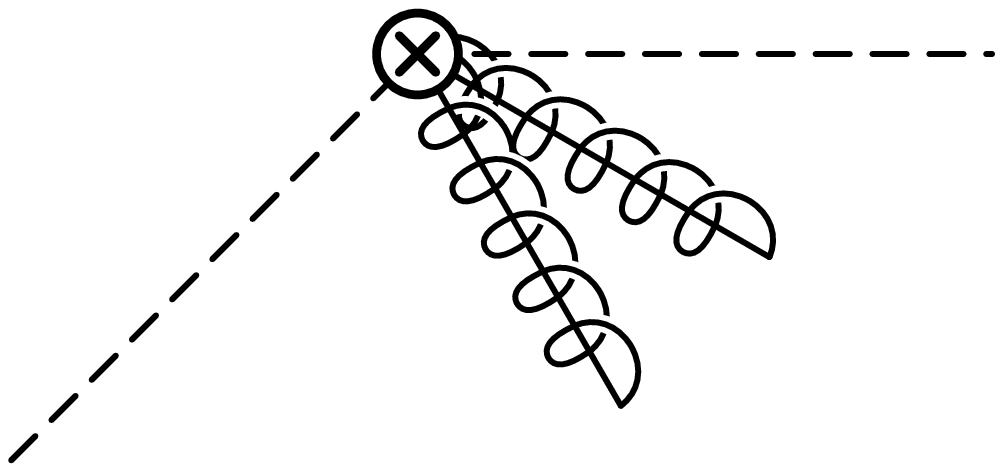}
\end{center}
\caption{\label{fig:wilson} Graphs in which $n$-collinear gauge bosons couple
to the $\bar n$-collinear line generate the Wilson line $W_n$. }
\end{figure}

Collinear gauge bosons and matter fields in the operator $\mathcal{O}$ occur in the combination $W_n^\dagger \xi_{n,p}$ or $W_n^\dagger \Phi_{n,p}$. They arise from full theory graphs in which the gauge fields couple to the \emph{other} particle, which moves in the $\bar n$-direction. The intermediate propagators are off-shell by order $Q^2$, and can be shrunk to a point, as shown in Fig.~\ref{fig:wilson}. 
Single gauge boson emission gives the vertex $g \bar n^\mu/(-\bar n \cdot k)$. At higher orders, multiple gauge boson emission from the $\bar n$ particle line combined with the non-Abelian multi-gluon interaction give the Wilson line operator $W_n$. Multi-gluon emission is related to single gluon emission by gauge invariance, and this relation holds even in the presence of loop-corrections. The Wilson line structure of the vertex $W_n$ is required by collinear gauge invariance~\cite{SCET}.

One has a similar construction for multiple $n$-collinear $\chi$ fields emitted from the $\bar n$ particle. At the level of single $\chi$ emission, the $\chi$ vertex is 
$h_{\psi,\phi}/(-\bar n \cdot k)$ instead of the gauge vertex $g \bar n^\mu/(-\bar n \cdot k)$. This is all we require for our computation. It would be interesting to work out the structure of the scalar vertex at higher orders, including radiative corrections. Multi-scalar emission is not related to the single scalar vertex by gauge invariance.

Most of the scalar corrections vanish. In SCET, the fermion Yukawa vertex vanishes, because Eq.~(\ref{constraint}) implies that
\begin{eqnarray}
\bar \xi_{n,p} \xi_{n,p} &=& \bar \xi_{n,p} \frac{\slashed{\bar n}\slashed{n}}{4}
\frac{\slashed{n}\slashed {\bar n}}{4} \xi_{n,p} =0
\end{eqnarray}
using $\slashed{n}\slashed{n}=n^2=0$. The tri-scalar couplings $\chi \phi^\dagger \phi $ have dimensions of mass, and $\chi$ exchange corrections to the scalar operators are suppressed by powers of $h_\phi/Q$, which is subleading in the EFT power counting given our assumption that $h_\phi$ does not grow with $Q$. The easiest way to see this is to use the rescaled fields $\phi_{n,p}$, which have a propagator of the same form as that for fermions. The Yukawa coupling becomes
\begin{eqnarray}
h_\phi\, \chi \phi^\dagger \phi &=& h_\phi\, \chi \Phi^\dagger_{n,p} \Phi_{n,p}= 
\frac{h_\phi}{\bar n \cdot p}\, \chi \phi^\dagger_{n,p} \phi_{n,p}\, ,
\end{eqnarray}
which is order $1/Q$ since $\bar n \cdot p$ is order $Q$, and gives an explicit $1/Q$ suppression to the graph at each tri-scalar vertex. 

One interesting point is the decoupling of scalars below $m_1$ and $m_2$, so that particles~1 and~2 can be treated as HQET fields. If one directly matches from the full theory onto HQET, then $h_\psi \chi \bar \psi \psi \to h_\psi \chi  \bar h_v h_v$, which is non-zero. If instead, one first goes through SCET, then $h_\psi \chi  \bar \psi \psi \to h_\psi \chi \bar \xi_{n,p} \xi_{n,p} =0$, so we have an apparent contradiction. However, the two results are not in disagreement. The scalar HQET vertex graph (Fig.~\ref{fig:dhqet} with the gauge boson replaced by a scalar) is equal to $-1/w$ times the corresponding gauge graphs, rescaled by the ratio of the Yukawa couplings to the gauge couplings. In the Sudakov limit, $w \sim Q^2/(m_1m_2)$, and $1/w$ is a power suppression which can be neglected, so both ways of matching agree, since power corrections are neglected.

\begin{table}
\begin{eqnarray*}
\begin{array}{|c|c|c|c|c|}
\hline
\text{Field} & m & M_\chi & \\[5pt]
\hline
\psi & m=0 & M_\chi=0 & \frac{1}{2\eUV}-\frac{1}{2\eIR} \\[5pt]
\psi & m=0 & M_\chi\not=0 &  \frac{1}{2\eUV}+\frac14 -\frac12\lchi \\[5pt]
\psi & m\not=0 & M_\chi=0 &  \frac{1}{2\eUV}-\frac{2}{\eIR}-\frac72+\frac32 \lm \\[5pt]
\psi & m\not=0 & M_\chi\not=0 &  \frac{1}{2\eUV}+\frac14 -\frac12\lchi + \tilde h_F(m^2/M_\chi^2) \\[5pt]
\phi & m=0 & M_\chi=0 & 0 \\[5pt]
\phi & m=0 & M_\chi\not=0 & \frac{1}{2M_\chi^2} \\[5pt]
\phi & m\not=0 & M_\chi=0 &  \frac{1}{m^2}\left[ -\frac{1}{2\eIR}-1
+\frac12 \lm \right] \\[5pt]
\phi & m\not=0 & M_\chi\not=0 &   \frac{1}{2M_\chi^2} + \frac{1}{M_\chi^2} \tilde h_S(m^2/M_\chi^2) \\[5pt]
h_v && M_\chi=0 & \frac{2}{\eUV}- \frac{2}{\eIR} \\[5pt]
h_v && M_\chi\not=0 &\frac{2}{\eUV}- 2 \lchi \\[5pt]
\hline
\end{array}
\end{eqnarray*}
\caption{\label{tab:swave} One-loop scalar exchange contribution to on-shell wavefunction renormalization. The exchanged scalar mass is $M_\chi$, $\lchi=\log M^2_\chi/\mu^2$, and the particle (fermion or scalar) mass is $m$. $\tilde h_{F,S}$ are given in Appendix~\ref{app:integrals}. An overall factor of $h_1 h_2/(16\pi^2)$ is omitted.}
\end{table}

The only scalar graphs which remain are the matching at $Q$, which are full theory graphs, and scalar contributions to wavefunction renormalization in the effective theories. The wavefunction contributions are summarized in Table~\ref{tab:swave}, where $\lchi=\log M^2_\chi/\mu^2$. The full theory wavefunction renormalization vanishes for both fermions and scalars, so the entire matching correction at $Q$ given in Eq.~(\ref{9.25}) arises from the vertex correction. The EFT matching and running can be computed from the scalar wavefunction graphs in Table~\ref{tab:swave}. The EFT matrix elements are given by taking $-1/2$ times the entries in the table for each particle, multiplying by $h^2/(16\pi^2)$, and then adding the contributions from the two particles. The finite part gives the matching correction, and $(-2)$ times the coefficient of the $1/\eUV$ term gives the anomalous dimension. Since only wavefunction graphs contribute, there are no $\lQ$ terms which can only arise from vertex graphs.

The computation of scalar contributions to the anomalous dimension and matching for the various cases considered in Sec.~\ref{sec:mass} parallels the gauge boson discussion. The matching coefficients for $m_{1,2}\not=0$ are given by a formula analogous to Eq.~(\ref{77}), with the gauge boson functions $f_{F,S}$ and $h_{F,S}$ replaced by the corresponding $\chi$-exchange functions $f_{F,S} \to 0$, since there are no vertex corrections in the effective theory,  and $h_{F,S} \to \tilde h_{F,S}$.

We have divided the scalar exchange contributions into vertex and wavefunction pieces, rather than giving the total contribution as in the gauge case. The reason is that the standard model is a chiral gauge theory, and the Yukawa couplings connect one matter representation to another. Thus, vertex corrections can mix left-handed currents with right-handed currents, whereas wavefunction corrections do not mix different $SU(2) \times U(1)$ representations. Keeping the two contributions separate allows us to compute Higgs radiative corrections in the standard model using the the results given in this section.

The matching corrections $C$ at $Q$ can be computed as for the gauge boson case. The matching corrections for external scalar particles due to scalar $\chi$ exchange are power suppressed, and vanish to leading order. For fermions, the $\chi$ exchange corrections give
\begin{eqnarray}
\bar \psi \psi &\to & \exp\left[ a_Y \left(-2 + \lQ\right)\right]  [\bar \xi_{n,p_2} W_n]
[W^\dagger_{\bar n} \xi_{\bar n,p_1}] \nn
\bar \psi \gamma^\mu \psi &\to & \exp\left[ a_Y \left(\frac12 -\frac12 \lQ\right)\right]  [\bar \xi_{n,p_2} W_n]\gamma^\mu [W^\dagger_{\bar n} \xi_{\bar n,p_1}] \nn
\bar \psi \sigma^{\mu \nu} \psi &\to & \exp\left[a_Y\right]\biggl\{[\bar \xi_{n,p_2} W_n]\sigma^{\mu \nu} [W^\dagger_{\bar n} \xi_{\bar n,p_1}]\nn
&&- \frac{i}{2}\left(n^\mu \bar n^\nu-n^\nu \bar n^\mu\right) [\bar \xi_{n,p_2} W_n][W^\dagger_{\bar n} \xi_{\bar n,p_1}]\biggr\} \nn
a_Y &=& \frac{h_{\psi,1} h_{\psi,2}}{16\pi^2}
\label{9.25}
\end{eqnarray}
which have to be combined with the gauge boson matching conditions in Table~\ref{tab:results}. The vertex graph contributes $2 a_Y$, $-a_Y$ and $0$, respectively, to the anomalous dimensions of the three operators in the full theory. The wavefunction graphs contribute an additional $a_Y$ to all three operators.

\section{Application to the Standard Model}
\label{sec:stdmodel}

The results we have obtained for the toy theory can now be used to compute results for the standard model.  One has to be careful in using the correct coupling constants, since the standard model is a chiral gauge theory, and the toy model is vector-like.

For the LHC, one is interested in processes such as dijet production. The SCET operators at the high scale $Q$ involve more than two SCET fields. E.g. in $q \bar q \to q \bar q$, the EFT operator has four fields, two for the incoming particle and two for the outgoing ones. One can obtain results for more than two external particles by combining the two-particle results computed in this paper with the appropriate gauge theory factors such as $C_A$ and $C_F$. There are several interesting features of the analysis which are independent of the calculations presented in this paper, so we defer the discussion of experimentally relevant examples to a subsequent publication~\cite{cgkm3}. Here we show how our results can be used to compute the radiative corrections to quark production by a gauge invariant current $\bar Q_i \gamma^\mu P_L Q_i$, where $Q_i$ is the quark doublet\footnote{Not to be confused with $Q$, the momentum transfer.} for generation $i=u,c,t$, and to charged lepton production by $\bar L \gamma^\mu P_L L$. We will do the computations for light quarks in Sec.~\ref{sec:up}, for leptons in Sec.~\ref{sec:leptons}, and for top quarks in Sec.~\ref{sec:top}. All fermion masses other than the top quark mass are neglected.

\subsection{Light Quarks}
\label{sec:up}

The  first generation quark doublet  is
\begin{eqnarray}
Q_u &=& \left( \begin{array}{cc} u \\ d^\prime \end{array}\right)=
\left( \begin{array}{cc} t \\ V_{ud}d + V_{us} s+ V_{ub} b \end{array}\right)\, ,
\label{doublet}
\end{eqnarray}
using the mass eigenstate basis. At the scale $Q \gg m_q$ the coefficient of the operator in the full electroweak theory is assumed to be unity. For the first generation, all quark masses and Yukawa couplings can be neglected, and so the answer is given by combining the gauge boson contributions computed earlier.

The operator in SCET at the scale $Q$ is
\begin{eqnarray}
\bar Q_u \gamma^\mu P_L Q_u &\to &c(Q) [\bar \xi_{n,p_2}^{(Q_u)} W_n] \gamma^\mu P_L
[W^\dagger_{\bar n} \xi^{(Q_u)}_{\bar n,p_1}]\, , \nn
\label{82}
\end{eqnarray}
where $\xi^{(Q_u)}$represents the left-handed electroweak $u$-quark doublet Eq.~(\ref{doublet}) in SCET, and we have suppressed gauge indices. The matching condition is\begin{eqnarray}
\log c(Q) &=& \left[ \frac{\alpha_s(Q)}{4 \pi}\frac{4}{3}+\frac{\alpha_2(Q)}{4 \pi}\frac{3}{4}+\frac{\alpha_1(Q)}{4 \pi}\frac{1}{36}\right] \left[ \frac{\pi^2}{6}-8\right]\, ,\nn
\label{83}
\end{eqnarray}
using the third column of  Table~\ref{tab:results} with $\lQ=0$ at the scale $\mu=Q$. The gauge couplings have been multiplied by the corresponding $C_F$ values: $4/3$ for an $SU(3)$ triplet, $3/4$ for an $SU(2)$ doublet, and $1/36$ for $Y=1/6$. The electroweak couplings renormalized at $\mu=M_Z$ are
\begin{eqnarray}
\alpha_2(M_Z) &=& \frac{\aem(M_Z)}{\sin^2 \theta_W(M_Z)}\, ,\nn
\alpha_1(M_Z) &=& \frac{\aem(M_Z)}{\cos^2 \theta_W(M_Z)}\, ,
\end{eqnarray}
and their values at $Q$ are obtained by the usual $\beta$-functions of the standard model. 

The theory below $Q$ is SCET with an $SU(3)\times SU(2) \times U(1)$ gauge symmetry. In this regime, the SCET current in Eq.~(\ref{82}) is multiplicatively renormalized with anomalous dimension (from the fourth column in Table~\ref{tab:results})
\begin{eqnarray}
\gamma(\mu) &=&\left[ \frac{\alpha_s(\mu)}{4 \pi}\frac{4}{3}+\frac{\alpha_2(\mu)}{4 \pi}\frac{3}{4}+\frac{\alpha_1(\mu)}{4 \pi}\frac{1}{36}\right] \left[ 4\lQ-6\right]\ .\nn
\label{85}
\end{eqnarray}
The anomalous dimension $\gamma$ is used to run $c$ down to a scale of order 
the gauge boson mass. One can integrate out the weak gauge bosons sequentially, by first integrating out the $Z$ boson at $\mu=M_Z$, followed by the $W$ at $\mu=M_W$.
This sums $\left(\alpha \log^2 M_W/M_Z\right)^n$, $n>1$ terms, while neglecting $\alpha \left(M_W/M_Z\right)^n$, $n>0$ power corrections. This is not a good choice to use for the standard model, since $M_W/M_Z$ is not very small, and summing powers of $M_W/M_Z$ is more important than summing $\alpha \log^2 M_W/M_Z$ terms. Instead, we integrate out the $W$ and $Z$ at a common scale, chosen to be $\mu=M_Z$. In this way, we match directly from an $SU(3) \times SU(2) \times U(1)$ gauge theory onto a $SU(3) \times U(1)_{\text{em}}$ gauge theory of gluons and photons, and there are no complications of an intermediate stage of broken electroweak symmetry where the $Z$ is integrated out, but not the $W$.

At the scale $\mu=M_Z$, integrating out the $W$ and $Z$ bosons give a matching correction to the SCET operator,
\begin{eqnarray}
 [\bar \xi_{n,p_2}^{(Q_u)} W_n] \gamma^\mu P_L
[W^\dagger_{\bar n} \xi^{(Q_u)}_{\bar n,p_1}] 
&\to& a^{(u)} [\bar \xi_{n,p_2}^{(u)} W_n] \gamma^\mu P_L
[W^\dagger_{\bar n} \xi^{(u)}_{\bar n,p_1}] \nn
&&\hspace{-3cm} + a^{(d^\prime)} [\bar \xi_{n,p_2}^{(d^\prime)} W_n] \gamma^\mu P_L
[W^\dagger_{\bar n} \xi^{(d^\prime)}_{\bar n,p_1}]\, .
\label{86}
\end{eqnarray}
Since the electroweak symmetry  is broken, the $u$ and $d^\prime$ parts of the operator get different matching corrections. The corrections $a_i$ are obtained using the last column of Table~\ref{tab:results}:
\begin{widetext}
\begin{eqnarray}
\log a^{(u)}(M_Z)&=&
\frac{\aem}{4\pi \sin^2 \theta_W \cos^2 \theta_W}
\left(\frac12-\frac23\sin^2\theta_W\right)^2 \Biggl[ \frac92-\frac{5\pi^2}{6}\Biggr]\nn
&&+\frac{\aem}{4\pi \sin^2 \theta_W}\left(\frac12\right)\Biggl[- \log^2 \frac{M_W^2}{M_Z^2}+2  \log \frac{M_W^2}{M_Z^2} \log \frac{Q^2}{M_Z^2}- 3 \log \frac{M_W^2}{M_Z^2}+ \frac92-\frac{5\pi^2}{6}\Biggr]\, ,\nn
\log a^{(d^\prime)}(M_Z)&=&
\frac{\aem}{4\pi \sin^2 \theta_W \cos^2 \theta_W}
\left(-\frac12+\frac13\sin^2\theta_W\right)^2 \Biggl[ \frac92-\frac{5\pi^2}{6}\Biggr]\nn
&&+\frac{\aem}{4\pi \sin^2 \theta_W}\left(\frac12\right)\Biggl[- \log^2 \frac{M_W^2}{M_Z^2}+2  \log \frac{M_W^2}{M_Z^2} \log \frac{Q^2}{M_Z^2}- 3 \log \frac{M_W^2}{M_Z^2}+ \frac92-\frac{5\pi^2}{6}\Biggr]\, .
\label{87}
\end{eqnarray}
\end{widetext}
The first term for $\log a^{(u,d^\prime)}$ is the $Z$ contribution, the second term is the $W$ contribution, and the coupling constants are renormalized at $M_Z$.

Below $M_Z$, the operators in Eq.~(\ref{86}) are multiplicatively renormalized, with anomalous dimensions
\begin{eqnarray}
\gamma^{(u)}&=&\left[ \frac{\alpha_s(\mu)}{4 \pi}\frac{4}{3}+\frac{\aem(\mu)}{4 \pi}\frac49\right] \left[ 4\lQ-6\right]\, ,\nn
\gamma^{(d^\prime)}&=&\left[ \frac{\alpha_s(\mu)}{4 \pi}\frac{4}{3}+\frac{\aem(\mu)}{4 \pi}\frac19\right] \left[ 4\lQ-6\right]\, ,
\label{88}
\end{eqnarray}
for the $u$ and $d^\prime$ terms.

The final result for the operator at a low scale is
\begin{eqnarray}
\bar Q_u \gamma^\mu P_L Q_u &\to& c^{(u)} [\bar \xi_{n,p_2}^{(u)} W_n] \gamma^\mu P_L
[W^\dagger_{\bar n} \xi^{(u)}_{\bar n,p_1}]\nn
&&\hspace{-3cm} + c^{(d^\prime)} [\bar \xi_{n,p_2}^{(d^\prime)} W_n] \gamma^\mu P_L
[W^\dagger_{\bar n} \xi^{(d^\prime)}_{\bar n,p_1}]\, ,
\label{103a}
\end{eqnarray}
with
\begin{eqnarray}
\log c^{(u)}(\mu) &=& \log c(Q) + \int_Q^{M_Z} \frac{ \rd \mu}{\mu} \gamma(\mu) \nn
&&+ \log a^{(u)} + \int_{M_Z}^{\mu} \frac{ \rd \mu}{\mu} \gamma^{(u)}(\mu)\, ,\nn
\log c^{(d^\prime)}(\mu) &=& \log c(Q) + \int_Q^{M_Z} \frac{ \rd \mu}{\mu} \gamma(\mu) \nn
&&+ \log a^{(d^\prime)} + \int_{M_Z}^{\mu} \frac{ \rd \mu}{\mu} \gamma^{(d^\prime)}(\mu)\, ,
\label{90}
\end{eqnarray}
where the various pieces are given in Eqs.~(\ref{83},\ref{85},\ref{87},\ref{88}). The EFT operator Eq.~(\ref{103a}) can then be used to compute processes such as dijet production using SCET~\cite{2jet}. For jet production, the scale $\mu$ would be chosen to be of order the jet invariant mass, around 30~GeV for jets at the LHC.

\subsection{Leptons}
\label{sec:leptons}

The computation for the radiative corrections to the lepton current $\bar L \gamma^\mu P_L L$, where $L$ is the lepton doublet
\begin{eqnarray}
L &=& \left( \begin{array}{cc} \nu \\ \ell \end{array}\right)\, ,
\label{ldoublet}
\end{eqnarray}
is similar to that for the quark doublet, and we summarize the final result. The full theory operator at the low scale $\mu$ is
\begin{eqnarray}
\bar L \gamma^\mu P_L L &\to& c^{(\nu)} [\bar \xi_{n,p_2}^{(\nu)} W_n] \gamma^\mu P_L
[W^\dagger_{\bar n} \xi^{(\nu)}_{\bar n,p_1}]\nn
&&\hspace{-3cm} + c^{(\ell)} [\bar \xi_{n,p_2}^{(\ell)} W_n] \gamma^\mu P_L
[W^\dagger_{\bar n} \xi^{(\ell)}_{\bar n,p_1}]\, , 
\end{eqnarray}
with the coefficients given by Eq.~(\ref{90}) with $u \to \nu$, $d^\prime \to \ell$, where the on the rhs of Eq.~(\ref{90}) for leptons are:
\begin{eqnarray}
\log c(Q) &=& \left[ \frac{\alpha_2(Q)}{4 \pi}\frac{3}{4}+\frac{\alpha_1(Q)}{4 \pi}\frac{1}{4}\right] \left[ \frac{\pi^2}{6}-8\right]\, ,\nn
\gamma(\mu) &=&\left[ \frac{\alpha_2(\mu)}{4 \pi}\frac{3}{4}+\frac{\alpha_1(\mu)}{4 \pi}\frac{1}{4}\right] \left[ 4\lQ-6\right]\, ,\nn
\gamma^{(\nu)}&=& 0\, , \nn
\gamma^{(\ell)}&=&\frac{\alpha(\mu)}{4 \pi}\left[ 4\lQ-6\right]\, ,
\label{l83}
\end{eqnarray}
\begin{widetext}
\begin{eqnarray}
\log a^{(\nu)}(M_Z)&=&
\frac{\aem}{4\pi \sin^2 \theta_W \cos^2 \theta_W}
\left(\frac12\right)^2 \Biggl[ \frac92-\frac{5\pi^2}{6}\Biggr]\nn
&&+\frac{\aem}{4\pi \sin^2 \theta_W}\left(\frac12\right)\Biggl[- \log^2 \frac{M_W^2}{M_Z^2}+2  \log \frac{M_W^2}{M_Z^2} \log \frac{Q^2}{M_Z^2}- 3 \log \frac{M_W^2}{M_Z^2}+ \frac92-\frac{5\pi^2}{6}\Biggr]\, ,\nn
\log a^{(\ell)}(M_Z)&=&
\frac{\aem}{4\pi \sin^2 \theta_W \cos^2 \theta_W}
\left(-\frac12+\sin^2\theta_W\right)^2 \Biggl[ \frac92-\frac{5\pi^2}{6}\Biggr]\nn
&&+\frac{\aem}{4\pi \sin^2 \theta_W}\left(\frac12\right)\Biggl[- \log^2 \frac{M_W^2}{M_Z^2}+2  \log \frac{M_W^2}{M_Z^2} \log \frac{Q^2}{M_Z^2}- 3 \log \frac{M_W^2}{M_Z^2}+ \frac92-\frac{5\pi^2}{6}\Biggr]\, .
\label{l87}
\end{eqnarray}
\end{widetext}

\subsection{Top Quarks}
\label{sec:top}

In this subsection, we show how our results can be used to compute the radiative corrections to  $t \bar t$ production by a gauge invariant vector current $\bar Q_t \gamma^\mu P_L Q_t$, where $Q_t$ is the left-handed quark doublet in the standard model,
\begin{eqnarray}
Q_t &=& \left( \begin{array}{cc} t \\ b^\prime \end{array}\right)=
\left( \begin{array}{cc} t \\ V_{td}d + V_{ts} s+ V_{tb} b \end{array}\right)\, ,
\label{tdoublet}
\end{eqnarray}
and $b^\prime=V_{td}d + V_{ts} s+ V_{tb} b$ using the mass eigenstate basis. We will neglect all quark masses other than $m_t$. This example illustrates how to use the fermion mass and Higgs exchange contributions computed in the toy example.

\begin{figure}
\begin{center}
\includegraphics[width=4cm]{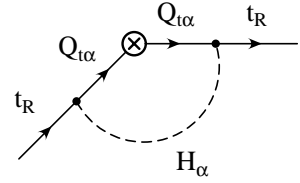}
\end{center}
\caption{\label{fig:higgs} Higgs correction which causes $\bar Q_t \gamma^\mu P_L Q_t$ to mix with $\bar t_R \gamma^\mu P_R t$. The index $\alpha$ is an $SU(2)$ index, and is summed over.}
\end{figure}

The operator in SCET at the scale $Q$ is
\begin{eqnarray}
\bar Q_t \gamma^\mu P_L Q_t &\to &c_1(Q) [\bar \xi_{n,p_2}^{(Q_t)} W_n] \gamma^\mu P_L
[W^\dagger_{\bar n} \xi^{(Q_t)}_{\bar n,p_1}] \nn
&&\hspace{-3cm}+c_2(Q) [\bar \xi_{n,p_2}^{(t)} W_n] \gamma^\mu P_R
[W^\dagger_{\bar n} \xi^{(t)}_{\bar n,p_1}] \, ,
\label{t82}
\end{eqnarray}
where $\xi^{(Q_t)}$ and $\xi^{(t)}$  represents the left-handed electroweak $t$-quark doublet Eq.~(\ref{tdoublet}) and the right-handed $t$-quark singlet $t_R$ in SCET and we have suppressed gauge indices. The $t_R$ terms arise from Higgs exchange graphs~Fig.~\ref{fig:higgs}.

The matching condition is
\begin{eqnarray}
\log c_1(Q) &=& \left[ \frac{\alpha_s(Q)}{4 \pi}\frac{4}{3}+\frac{\alpha_2(Q)}{4 \pi}\frac{3}{4}+\frac{\alpha_1(Q)}{4 \pi}\frac{1}{36}\right] \left[ \frac{\pi^2}{6}-8\right]\, ,\nn
c_2(Q)&=&\left[2 \frac{g_t^2(Q)}{16\pi^2}\right]\left[ \frac{1}{2}\right]\, ,
\end{eqnarray}
using Tables~\ref{tab:results} and Eq.~(\ref{9.25}) with $\lQ=0$ at the scale $\mu=Q$. The gauge couplings have been multiplied by the corresponding $C_F$ values, $4/3$ for an $SU(3)$ triplet, $3/4$ for an $SU(2)$ doublet, and $1/36$ for $Y=1/6$.  The top quark Yukawa coupling is normalized so that $g_t=\sqrt 2 m_t/v$, with $v\sim 247$~GeV. The Higgs exchange graph in the chiral standard model has been computed using the toy-model value Eq.~(\ref{9.25}) for a vector-like theory, combined with the result that a Yukawa vertex flips the fermion chirality. The factor of 2 in front of $g_t^2/(16\pi^2)$ arises from summing over a  closed $SU(2)$ index loop, i.e.\ because the Higgs and $Q$ are $SU(2)$ doublets (the sum on $\alpha$ in Fig.~\ref{fig:higgs}). The factor of $1/2$ in square brackets is the coefficient of $a_Y$ in the second line of Eq.~(\ref{9.25}), with $\lQ \to 0$ at $\mu=Q$. The Higgs exchange vertex correction mixes the $Q_L$ operator with the $t_R$ operator. Higgs exchange corrections do not contribute to the diagonal coefficient $c_1$, since the first row of Table~\ref{tab:swave} shows that the full theory wavefunction renormalization has no finite part.

The theory below $Q$ is SCET with an $SU(3)\times SU(2) \times U(1)$ gauge symmetry. In this regime, the two operators in Eq.~(\ref{t82}) are multiplicatively renormalized with anomalous dimensions
\begin{eqnarray}
\mu \frac{\rd c_1}{\rm d \mu}  &=&\Biggl\{ \left[ \frac{\alpha_s(\mu)}{4 \pi}\frac{4}{3}+\frac{\alpha_2(\mu)}{4 \pi}\frac{3}{4}+\frac{\alpha_1(\mu)}{4 \pi}\frac{1}{36}\right] \left[ 4\lQ-6\right]\nn
&&+\frac{g_t^2(Q)}{16\pi^2}\Biggr\}c_1\, ,\nn
\mu \frac{\rd c_2}{\rm d \mu}  &=&\Biggl\{ \left[ \frac{\alpha_s(\mu)}{4 \pi}\frac{4}{3}+\frac{\alpha_1(\mu)}{4 \pi}\frac{4}{9}\right] \left[ 4\lQ-6\right]\nn
&&+2 \frac{g_t^2(Q)}{16\pi^2}\Biggr\}c_2\, .
\label{t85}
\end{eqnarray}
The $t_L$ wavefunction factor due to Higgs exchange does not have the factor of two from the $SU(2)$ index summation that is present for $t_R$. The Higgs vertex graph, which causes $c_1-c_2$ mixing, is $1/Q^2$ suppressed.

The anomalous dimension $\gamma$ is used to run $c_{1,2}$ down to a scale of order $m_t$. At this scale there are several different methods one can use. As for massless quarks, one can integrate out the scales $m_t$, $M_W$, $M_Z$ and $M_H$ in various ways, e.g.\ one can integrate out each particle at a scale $\mu$ equal to its mass, or integrate out one or more particles simultaneously at some common value of $\mu$. Integrating out the top quark leads to a complicated effective theory with dynamical $W$ and $Z$ bosons which is no longer $SU(2)\times U(1)$ invariant, since the $b^\prime$ quark is in the theory but not $t$. Luckily, the best method for experimentally relevant computations is also the simplest to use: Since $m_t$, $M_W$, $M_Z$, and presumably $M_H$ are not widely separated, one can integrate them all out together. In this way, one goes directly from an $SU(3) \times SU(2) \times U(1)$ invariant theory to a $SU(3) \times U(1)_{\text{em}}$ gauge theory,  with broken $SU(2) \times U(1)$ symmetry and no electroweak gauge bosons. This procedure keeps the entire mass dependence on the four mass scales.

At the scale $\mu=m_t$ the $t$-quark SCET field is replaced by the heavy quark field $t_v$, whereas the $b^\prime$ quark SCET field in the doublet $\xi^{(Q_t)}$ remains an SCET field $\xi^{(b^\prime)}$.
The operator matching is
\begin{eqnarray}
[\bar \xi_{n,p_2}^{(Q_t)} W_n] \gamma^\mu P_L [W^\dagger_{\bar n} \xi^{(Q_t)}_{\bar n,p_1}]  &\to& \frac12 a_1\bar t_{v_2} t_{v_1} \nn
&&\hspace{-3cm} + a_2 [\bar \xi_{n,p_2}^{(b^\prime)} W_n] \gamma^\mu P_L [W^\dagger_{\bar n} \xi^{(b^\prime)}_{\bar n,p_1}] \, ,\nn
{}[\bar \xi_{n,p_2}^{(t)} W_n] \gamma^\mu P_R [W^\dagger_{\bar n} \xi^{(t)}_{\bar n,p_1}]  &\to& \frac12 a_3 \bar t_{v_2} t_{v_1} \, ,
\label{t86}
\end{eqnarray}
where the matching coefficients are denoted $a_{1,2,3}$. Using the result of Sec.~\ref{sec:massG} for the gauge boson exchange graphs, and Sec.~\ref{sec:scalar} for the Higgs exchange graphs gives
\begin{widetext}
\begin{eqnarray}
F_g(Q,M,m) &=&  - \log^2 \frac{M^2}{m^2}+2  \log \frac{M^2}{m^2} \log \frac{Q^2}{m^2}- 3 \log \frac{M^2}{m^2} + \frac92-\frac{5\pi^2}{6}+2 f_F\left(\frac{m^2}{M^2}\right)-h_F\left(\frac{m^2}{M^2}\right)\nn
F_h(M,m)&=&\frac14-\frac12\log\frac{M^2}{m^2}+\tilde h_F\left(\frac{m^2}{M^2}\right)\nn
\log a_1(m_t)&=&
\frac{\aem}{4\pi \sin^2 \theta_W \cos^2 \theta_W}\left(\frac12-\frac23\sin^2\theta_W\right)^2 F_g(Q,M_Z,m_t)+\frac{\aem}{4\pi \sin^2 \theta_W}\left(\frac12\right)F_g(Q,M_W,m_t)\nn
&&+ \left(\frac{\alpha_s}{4\pi}\frac{4}{3}+\frac{\aem}{4\pi}\frac{4}{9}\right)\left(\frac{\pi^2}{6}+4\right)
-\left( \frac{g_t^2}{16\pi^2}\frac12\right)F_h(M_H,m_t)-\left(\frac{g_t^2}{16\pi^2}\frac12\right)F_h(M_Z,m_t)\, ,\nn
\log a_2(m_t)&=&
\frac{\aem}{4\pi \sin^2 \theta_W \cos^2 \theta_W}
\left(-\frac12+\frac13\sin^2\theta_W\right)^2 F_g(Q,M_Z,m_t)+\frac{\aem}{4\pi \sin^2 \theta_W}\left(\frac12\right)F_g(Q,M_W,m_t)\nn
&&+ \left(\frac{\alpha_s}{4\pi}\frac{4}{3}+\frac{\aem}{4\pi}\frac{1}{9}\right)\left(\frac{\pi^2}{6}+4\right)-\left(\frac{g_t^2}{16\pi^2}\right)F_h(M_W,m_t)\, ,\nn
\log a_3(m_t)&=&
\frac{\aem}{4\pi \sin^2 \theta_W \cos^2 \theta_W}
\left(-\frac23\sin^2\theta_W\right)^2F_g(Q,M_Z,m_t)+  \left(\frac{\alpha_s}{4\pi}\frac{4}{3}+\frac{\aem}{4\pi}\frac{4}{9}\right)\left(\frac{\pi^2}{6}+4\right)\nn
&&-\left(\frac{g_t^2}{16\pi^2}\frac12\right)F_h(M_H,m_t)-\left(\frac{g_t^2}{16\pi^2}\frac12\right)F_h(M_Z,m_t)
-\left(\frac{g_t^2}{16\pi^2}\right)F_h(M_W,m_t)\, .
\end{eqnarray}
\end{widetext}
All running couplings are renormalized at $\mu=m_t$. The expressions are given by adding the contributions due to the $Z$ ($F_g(Q,M_Z,m_t)$ term), $W$ ($F_g(Q,M_W,m_t)$ term), gluon, $\gamma$, $H$ ($F_h(M_H,m_t)$ term), $h^0$ ($F_h(M_Z,m_t)$ term) and $h^+$ ($F_h(M_W,m_t)$ term), where $h^0$, $h^+$ are the unphysical Higgs scalars present in $R_{\xi=1}$ gauge.

Below $\mu=m_t$, the $\bar t_{v_2} t_{v_1}$ operator has anomalous dimension (from
the third column of Table~\ref{tab:mmresults})
\begin{eqnarray}
\gamma_3 &=&\left[\frac{\alpha_s}{4\pi}\frac{4}{3}+\frac{\aem}{4\pi}\frac{4}{9}\right] 4\left[w r(w)-1\right]
\end{eqnarray}
where $w=v_2 \cdot v_1=1+ Q^2/(2m_t^2)$.

The radiative corrections to the $\bar t t$ operator can then be combined with known methods to obtain $t$-quark decay distributions~\cite{fhms}. The QCD corrections (the $\alpha_s$ terms) have already been included in the analysis of Ref.~\cite{fhms}. The new results in this paper are the additional electroweak radiative corrections, including Higgs effects.

\section{Numerics and Conclusions}

We have shown how SCET methods can be used to compute the radiative corrections to electroweak processes. We discussed the results for massless external particles given in Ref.~\cite{cgkm1} in more detail, and derived  the result that there is at most a single power of $\lQ$ in the matching at $M$ to all orders in perturbation theory. The existence of $\lQ$ terms in the matching at $M$ is a new feature of SCET with massive gauge bosons. The results of Ref.~\cite{cgkm1} have been extended to include external particle masses and radiative Higgs exchange corrections proportional to the Yukawa couplings. 

Most of the paper used the vector-like $SU(2)$ gauge theory. Section~\ref{sec:stdmodel} 
explained in detail how the the results for the vector-like theory could be used to compute radiative corrections in the standard model, which is a chiral gauge theory. In this paper, we have computed radiative corrections to operators with two external particles. These can be used to compute the production rate for two external particles by a gauge invariant bilinear source. This could be applied to the decay rate of a (hypothetical) gauge singlet particle into two fermions. SCET methods can also be used to compute the radiative corrections to the dominant high-energy processes observable at the LHC, such as dijet production from quark-quark scattering. For a partonic process such as $q q \to q q$, the EFT operator is a four-quark operator, and the radiative corrections can be obtained by using the results of this paper, summed over all pairs of particles. The anomalous dimensions are about twice as big as the ones for the two-quark operators considered here. We postpone further discussion to a subsequent publication~\cite{cgkm3}.

\begin{figure}
\begin{center}
\includegraphics[bb=18 156 592 542,width=9cm]{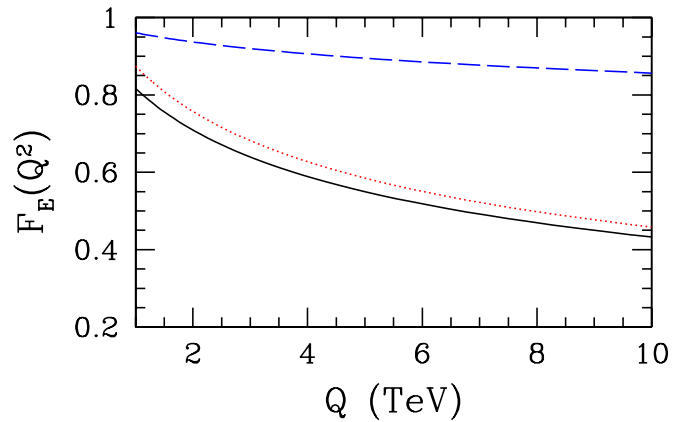}
\end{center}
\caption{\label{fig:p1} The Sudakov form-factor for $u$-quarks (solid black), $t$-quarks (dotted red) and electrons (dashed blue) at $\mu=M_Z$ for $m_H=200$~GeV. }
\end{figure}

\begin{figure}
\begin{center}
\includegraphics[bb=18 156 592 542,width=9cm]{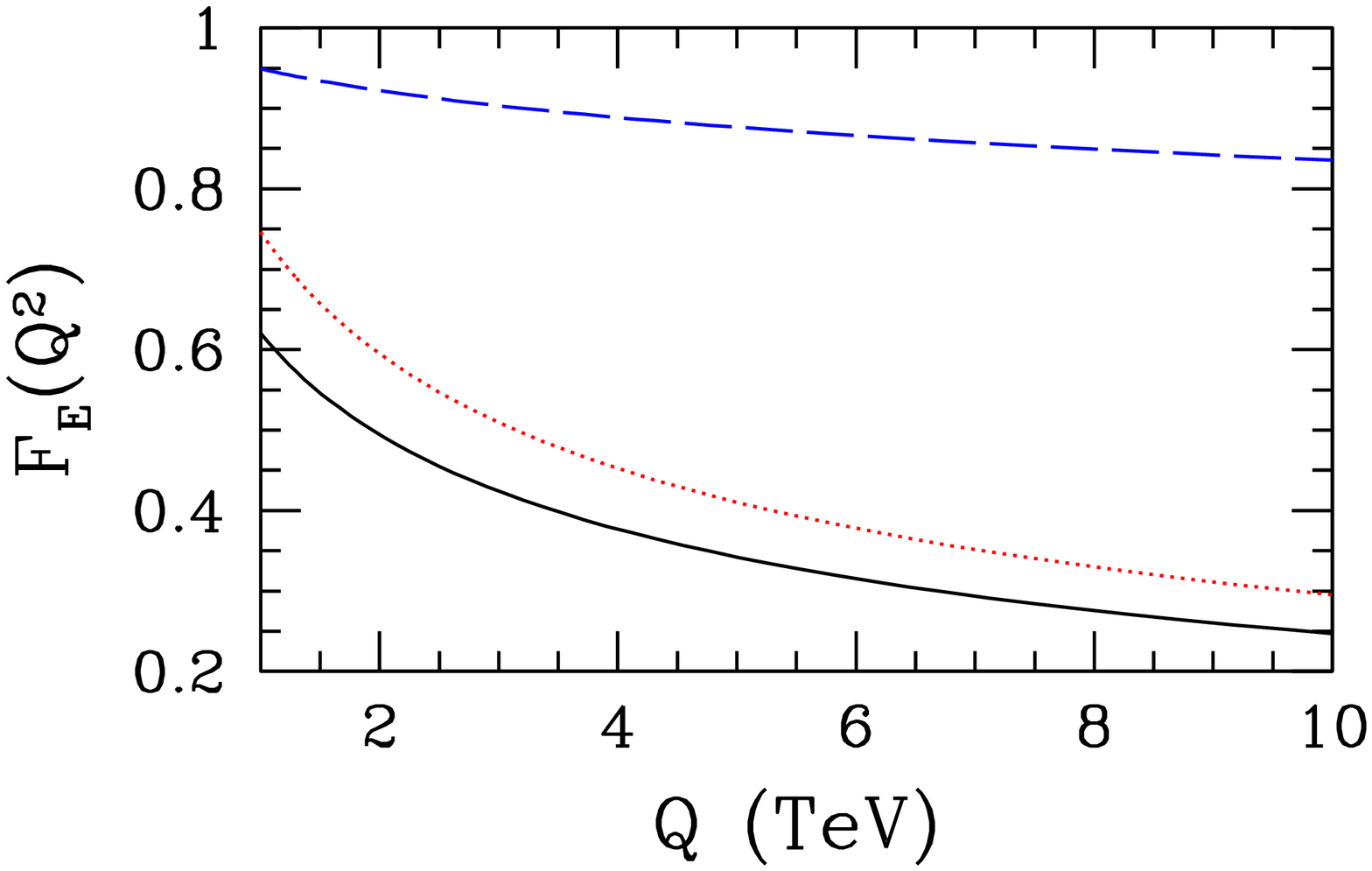}
\end{center}
\caption{\label{fig:p2} The Sudakov form-factor for $u$-quarks (solid black), $t$-quarks (dotted red) and electrons (dashed blue) at $\mu=30$~GeV for $m_H=200$~GeV. }
\end{figure}

\begin{figure}
\begin{center}
\includegraphics[bb=18 156 592 542,width=9cm]{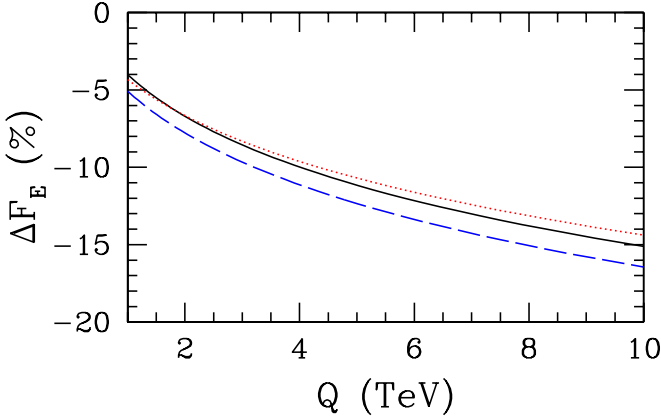}
\end{center}
\caption{\label{fig:p3} The electroweak contribution to the Sudakov form-factor (as a percentage change) for $u$-quarks (solid black), $t$-quarks (dotted red) and electrons (dashed blue) at $\mu=30$~GeV for $m_H=200$~GeV. }
\end{figure}

We conclude by giving some plots which show the typical size of the radiative corrections. The LHC center of mass energy is $\sqrt s =14$~TeV. The partonic center of mass energy $\sqrt{ \hat s}$ is much lower, because a proton with energy $E$ has partons with energy fraction $x \le 1$, given by  a parton distribution $f(x)$ which vanishes as $x \to 1$. The bulk of the dijet cross-section is at low $\hat s$, but the LHC has sufficient luminosity to be able to observe the dijet cross-section up to  $\sqrt{ \hat s}$ of order several TeV. We plot the Sudakov form factor for electron production via $\bar L \gamma^\mu P_L L$, $u$-quark production via $\bar Q_u \gamma^\mu P_L Q_u$ and $t$-quark production via $\bar Q_t \gamma^\mu P_L Q_t$, which are the results given in Sec.~\ref{sec:stdmodel}, for $\sqrt{\hat s}$ between $0.25$ and 8~TeV. Figure~\ref{fig:p1} gives the results for $F_E(Q^2)$ for the three cases, stopping the evolution at $\mu=M_Z$. In Fig.~\ref{fig:p2}, the EFT operators have been evolved all the way down to $\mu=30$~GeV, the typical invariant mass used to define a jet at the LHC.  Figure~\ref{fig:p3} shows the electroweak contributions to the three form-factors as a percentage change relative to the form factor including only the QCD radiative corrections. We have used a Higgs mass of 200~GeV in the plots. Varying the Higgs mass between 150~GeV and 500~GeV makes a difference of less than 0.5\%. 
$F_E$ is normalized to unity in the absence of radiative corrections. Radiative corrections for electrons are about 7\% at 2~TeV, and $\mu=M_Z$, increasing to about 8\% for $\mu=30$~GeV due to the QED running below $M_Z$. The corrections for quarks are much larger, 15\% for the $t$-quark, and 30\% for the $u$-quark at $\mu=M_Z$, increasing to 40\% and 50\%, respectively, at $\mu=30$~GeV. There is also a significant difference between the results for $t$- and $u$-quarks, arising from power corrections which depend on $m_t/M_{W,Z}$ and Higgs corrections which depend on the Yukawa coupling $g_t$. The bulk of the difference is due to the power corrections, the $g_t$ terms are less than 1\%.\footnote{Note that the Higgs corrections do not have the $\lQ$ enhancement that is present for the gauge bosons.} 
The corrections for the four-quark operators needed for realistic processes are bigger, with corrections even to color singlet processes being greater than 20\%. The radiative corrections are large enough that resummation is necessary to get an accurate prediction for the partonic cross-sections. We have shown how one can perform the resummation using EFT methods. Previous computations~\cite{ciafaloni,fadin,kps,fkps,jkps,jkps4,beccaria,dp1,dp2,hori,beenakker,dmp,pozzorini,js} have done the resummation using infrared evolution equations~\cite{fadin}. The EFT method allows one to include mass effects as well as Higgs corrections in a systematic way, which have not been included previously.  It also handles the cross-over between the $SU(3)\times SU(2)\times U(1)$ and the $SU(3) \times U(1)$ gauge theories above and below the weak interaction symmetry breaking scale, including the effects of unequal $W$ and $Z$ masses. The infrared evolution method uses a conjectured form for this cross-over with equal $W$ and $Z$ masses. The extension of previous results to massive external particles is currently being studied by other groups using a method-of-regions analysis~\cite{jantzentalk}.

RK was supported by an LHC theory fellowship from the NSF.

\begin{appendix}

\section{Analytic Regulator Dependence of Collinear Graphs}\label{app:analytic}

The $n$-collinear contribution Eq.~(\ref{26}) can be written as
\begin{eqnarray}
I_n&=&\frac{\alpha_s}{4\pi}C_F c(\mu) \gamma^\mu  \Biggl[\frac{2\eta}{\delta \epsilon}
+\frac{2\eta}{\delta} \log \frac{\mu^2}{M^2}+\frac{1-\rho}{\epsilon^2}\nn
&& +\frac{1}{\epsilon}\biggl(2 + (1+\rho) \log \frac{\nu_1^-}{p_2^-}
-(1-\rho) \log \frac{\nu_2^2}{\mu^2} \biggr) \nn
 &&+2+2 \log \frac{\mu^2}{M^2}-(1-\rho)\log \frac{\mu^2}{M^2}
 \log \frac{\nu_2^2}{\mu^2}\nn
 &&+(1+\rho) \log \frac{\mu^2}{M^2}
 \log \frac{\nu_1^-}{p_2^-}-\frac12(1-\rho) \log^2 \frac{\mu^2}{M^2}\nn
&& +\frac{\rho \pi^2}{12}-\frac{5 \pi^2}{12}
\Biggr]\, ,
\label{44}
\end{eqnarray}
where
\begin{eqnarray}
\eta = \frac{1}{r_1-r_2}\,, \qquad \rho=\frac{r_1+r_2}{r_1-r_2}\, .
\end{eqnarray}
The $\bar n$-collinear contribution is given by Eq.~(\ref{44}) with $\eta \to - \eta$, $\rho \to - \rho$, $\nu_2 \to \nu_1$, $\nu_1^-/p_2^- \to \nu_2^+/p_1^+$.  The dependence on $r_{1,2}$ and $\nu_{1,2}$ is additional scheme dependence introduced by the analytic regulator.  The sum of the $n$-collinear and $\bar n$-collinear contributions simplifies greatly, and gives Eq.~(\ref{28}) on using $\nu_1^- = \nu_1^2/p_1^+$, $\nu_2^+=\nu_2^2/p_2^-$. In particular, the $\eta$ and $\rho$ terms cancel between the two contributions. 
The relevant relations are
\begin{eqnarray}
\left[\log \frac{\nu_1^-}{p_2^-} + \log \frac{\nu_2^2}{\mu^2}\right]-\left[\log \frac{\nu_2^+}{p_1^+} + \log \frac{\nu_1^2}{\mu^2}\right] &=& \log \frac{\nu_1^- \nu_2^2 p_1^+ }{\nu_2^+ \nu_1^2 p_2^- }\nn
&=& 0\, , \nn
\left[\log \frac{\nu_1^-}{p_2^-} - \log \frac{\nu_2^2}{\mu^2}\right]+\left[\log \frac{\nu_2^+}{p_1^+} - \log \frac{\nu_1^2}{\mu^2}\right] &=& \log \frac{\nu_1^- \nu_2^+ \mu^4}{p_1^+ p_2^- \nu_1^2 \nu_2^2 } \nn
&=&-2\lQ\, .
\label{53}
\end{eqnarray}
The identities Eq.~(\ref{53}) do not make any assumptions about the values of $\nu_1^2$ and $\nu_2^2$, which need not be equal.

It is convenient to use the special case of Eq.~(\ref{44}) with $\rho=\eta=0$ to define the $n,\bar n$-collinear contributions. This form is given by using the analytic regulator followed by the limit $r_2=-r_1$, $r_1 \to \infty$, and gives
\begin{eqnarray}
I_n&=&\frac{\alpha_s}{4\pi}C_F c(\mu) \gamma^\mu  \Biggl[\frac{1}{\epsilon^2} +\frac{1}{\epsilon}\biggl(2 +  \log \frac{\nu_1^-}{p_2^-}
-\log \frac{\nu_2^2}{\mu^2} \biggr) \nn
 &&+2+2 \log \frac{\mu^2}{M^2}-\log \frac{\mu^2}{M^2}
 \log \frac{\nu_2^2}{\mu^2}\nn
 &&+ \log \frac{\mu^2}{M^2}
 \log \frac{\nu_1^-}{p_2^-}-\frac12 \log^2 \frac{\mu^2}{M^2} -\frac{5 \pi^2}{12}
\Biggr]\, ,
\label{45}
\end{eqnarray}
which has  no $1/\delta$ singularities. This form is similar to the value obtained in Ref.~\cite{zerobin} using a rapidity regulator.

\section{Parameter Integrals}
\label{app:integrals}

\subsection{Fermions}
\begin{eqnarray}
f_F(z) &=& 2 \int_0^{1} {\rd x }\frac{1-x}{x}\log\left(\frac{1-x+z x^2}{1-x}\right) \nn
\label{fF}
  &=& 2+\left(\frac{1}{z}-2\right) \log(z)  + \frac{2\sqrt{1-4z}}{z} \tanh^{-1}\sqrt{1-4z}\nn
&&+\frac12 \log^2 (z) - 2\left(\tanh^{-1} \sqrt{1-4z}\right)^2
\end{eqnarray}
\begin{eqnarray}
h_F(z) &=& -\int_0^1{\rm d}x\Biggl\{ 2(1-x) \log\left(\frac{1-x+z x^2}{1-x}\right)\nn
&&+\frac{4 z x(1-x^2)}{1-x+z x^2}\Biggr\} \nn
\label{hF}
 &=& \frac{9}{2} + \frac{3}{z} +  \left[ \frac{3}{2 z^2}-3 \right] \log(z) \nn
    &&+\frac{ \left(3-6z-12 z^2 \right)}{z^2 \sqrt{1-4 z}}   \tanh^{-1}(\sqrt{1-4 z})
    \end{eqnarray}
\begin{eqnarray} 
\tilde h_F(z) 
 &=&  -\int_0^{1}\rd x\  \Biggl\{ (1-x) \log\left( \frac{1-x+x^2 z}{1-x} \right) \nn
 &&-  \frac{2z x (1-x)(2-x)}{1-x+z x^2}\Biggr\} \nn
        &=&-\frac{15}{4}+\frac{3}{2 z}-\left[ \frac{3}{z}-\frac{3}{4 z^2}-\frac{3}{2} \right] \log(z)  \nn
    &&+\left[ \frac{ \sqrt{1-4 z} \left(3-6 z\right)}{2 z^2} \right] \tanh^{-1}(\sqrt{1-4z})\nn
    \end{eqnarray}
\subsection{Scalars}
\begin{eqnarray}
f_S(z) &=& \int_0^{1} {\rd x }  \frac{(2-x)}{x}  \log\frac{ 1-x+z x^2}{1-x} \nn
\label{fS}
   &=&1 -\left( 1-\frac{1}{2z} \right) \log(z) \nn
  && +\frac{\sqrt{1-4z} }{z} \tanh^{-1}(\sqrt{1-4z})  \nn
  &&+\frac12 \log^2 z - 2\left(\tanh^{-1} \sqrt{1-4z}\right)^2 
\end{eqnarray}
\begin{eqnarray}
h_S(z) &=&  \int_0^1{\rm d}x \Biggl\{(3x^2-6x+4)\log\left(\frac{1-x+z x^2}{1-x}\right)  \nn
&&\qquad-\frac{z x(1-x)(2-x)^2}{1-x+z x^2}  \Biggr\} \nn
\label{hS}
  &=& \frac{3}{2}-\frac{1}{z} + \left[ \frac{3}{2 z}-\frac{1}{2 z^2} \right] \log(z)\nn
 &&+ \left[ \frac{\sqrt{1-4 z}}{z^2} \left(z-1\right) \right] \tanh^{-1}(\sqrt{1-4 z})
\end{eqnarray}
\begin{eqnarray}
\tilde h_S(z) &=& = - \int_0^1{\rm d}x \frac{z x^3}{1-x+z x^2}  \nn
\label{htS}
 &=& -\frac{1}{2}-\frac{1}{z}  + \left[\frac{1}{2 z}-\frac{1}{2z^2}\right] \log(z)\nn
 && + \left[\frac{3z-1}{z^2\sqrt{1-4 z}} \right]   \tanh^{-1}(\sqrt{1-4z})  
\end{eqnarray}
For $4z \ge 1$, the results can be analytically continued using $\sqrt{1-4 z} \to i\sqrt{4 z-1}$ and $\tanh^{-1}(\sqrt{1-4 z}) \to i\tan^{-1}(\sqrt{4 z-1})$.  In each integral, the factors of $i$ cancel,  and the function remains real.

\end{appendix}

\end{document}